%% file: main.tex
\documentclass[sigconf]{acmart}

\AtBeginDocument{%
  }

\setcopyright{none}

\usepackage{booktabs}   %
\usepackage{subcaption} %
\usepackage{xspace}
\usepackage{makecell}
\usepackage{algorithm}
\usepackage[noend]{algorithmic}
\usepackage{tikz}
\usepackage{enumitem}
\input{macro}
\begin{document}

\title{Revisiting the Plastic Surgery Hypothesis \\ via Large Language Models}

\author{Chunqiu Steven Xia}
    \affiliation{\institution{University of Illinois Urbana-Champaign}\country{}}
    \email{chunqiu2@illinois.edu}
\author{Yifeng Ding}
    \affiliation{\institution{University of Illinois Urbana-Champaign}\country{}}
    \email{yifeng6@illinois.edu}
\author{Lingming Zhang}
    \affiliation{\institution{University of Illinois Urbana-Champaign}\country{}}
    \email{lingming@illinois.edu}

\begin{abstract}
\aprfull (\apr) aspires to automatically generate patches for an input buggy program. Traditional \apr tools typically focus on  specific bug types and fixes through the use of templates, heuristics, and formal specifications. However, these techniques are limited in terms of the bug types and patch variety they can produce. As such, researchers have designed various \learning \apr tools with recent work focused on directly using \plmfull{s} (\plm{s}) for \apr. While \llm-based \apr tools are able to achieve state-of-the-art performance on many repair datasets, the \plm{s} used for direct repair are not fully aware of the project-specific information such as unique variable or method names.

The \emph{plastic surgery hypothesis} is a well-known insight for \apr, which states that the code ingredients to fix the bug usually already exist within the same project. Traditional \apr tools have largely leveraged the plastic surgery hypothesis by designing manual or heuristic-based approaches to exploit such existing code ingredients. However, as recent \apr research starts focusing on \llm-based approaches, the plastic surgery hypothesis has been largely ignored. In this paper, we ask the following question: \emph{How useful is the plastic surgery hypothesis in the era of \plm{s}?}
Interestingly, \llm-based \apr presents a unique opportunity to fully automate the plastic surgery hypothesis via fine-tuning (training on the buggy project) and prompting (directly providing valuable code ingredients as hints to the \llm). To this end, we propose \ourtech, which combines the direct usage of \plm{s} with two domain-specific fine-tuning strategies and one prompting strategy (via information retrieval and static analysis) for more powerful \apr.%
While traditional \apr techniques require intensive manual efforts in both generating patches based on the plastic surgery hypothesis and guaranteeing patch validity, our approach is fully automated and general.%
Moreover, while it is very challenging to manually design heuristics/patterns for effectively leveraging the hypothesis, due to the power of \llm{s} in code vectorization/understanding, even partial/imprecise project-specific information can still guide \plm{s} in generating correct patches!
Our experiments on the widely studied \dfj 1.2 and 2.0 datasets show that \ourtech fixes 89 and 44 bugs (substantially outperforming the best-performing baseline by 15 and 8), respectively, demonstrating a promising future of the plastic surgery hypothesis in the era of \llm{s}.

\end{abstract}

\maketitle

\input{intro}

\input{background}

\input{approach}

\input{methodology}
\input{results}

\input{threat_to_validity}
\input{conclusion}

\bibliographystyle{ACM-Reference-Format}
\bibliography{reference.bib}

\end{document}

%% file: macro.tex
\newcommand{\CodeIn}[1]{{\small \texttt{#1}}}

\newcommand{\Comment}[1]{}
\newcommand{\mypara}[1]{\vspace{.03in}\noindent \textbf{#1}}

\newcommand{\apr}{APR\xspace}
\newcommand{\aprfull}{Automated Program Repair\xspace}
\newcommand{\plm}{LLM\xspace}
\newcommand{\plmfull}{Large Language Model\xspace}
\newcommand{\llm}{LLM\xspace}

\newcommand{\msp}{MSP\xspace}
\newcommand{\mspfull}{Masked Span Prediction\xspace}
\newcommand{\mlm}{MLM\xspace}
\newcommand{\mlmfull}{Masked Language Model\xspace}

\newcommand{\nmt}{NMT\xspace}
\newcommand{\nmtfull}{Neural Machine Translation\xspace}
\newcommand{\csapr}{cloze-style APR\xspace}
\newcommand{\Csapr}{Cloze-style APR\xspace}
\newcommand{\template}{template-based\xspace}
\newcommand{\Template}{Template-based\xspace}
\newcommand{\learning}{learning-based\xspace}

\newcommand{\ourtech}{FitRepair\xspace} %
\newcommand{\epfinetune}{Knowledge-Intensified fine-tuning\xspace}
\newcommand{\rofinetune}{Repair-Oriented fine-tuning\xspace}
\newcommand{\idprompting}{Relevant-Identifier prompting\xspace}

\newcommand{\ctfiveseed}{CodeT5$\times$4\xspace}

\newcommand{\ctfive}{CodeT5\xspace}
\newcommand{\codebert}{CodeBERT\xspace}

\newcommand{\dfj}{Defects4j\xspace}

\newcommand{\alpharepair}{AlphaRepair\xspace}
\newcommand{\selfapr}{SelfAPR\xspace}
\newcommand{\rewardrepair}{RewardRepair\xspace}
\newcommand{\recoder}{Recoder\xspace}

\newcommand{\cure}{CURE\xspace}
\newcommand{\coconut}{CoCoNuT\xspace}
\newcommand{\dlfix}{DLFix\xspace}
\newcommand{\sequencer}{SequenceR\xspace}
\newcommand{\tbar}{TBar\xspace}
\newcommand{\prapr}{PraPR\xspace}
\newcommand{\avatar}{AVATAR\xspace}
\newcommand{\simfix}{SimFix\xspace}
\newcommand{\fixminer}{FixMiner\xspace}
\newcommand{\capgen}{CapGen\xspace}
\newcommand{\jaid}{JAID\xspace}
\newcommand{\sketchfix}{SketchFix\xspace}
\newcommand{\nopol}{NOPOL\xspace}
\newcommand{\jgenprog}{jGenProg\xspace}
\newcommand{\jmutrepair}{jMutRepair\xspace}
\newcommand{\jkali}{jKali\xspace}

\newcommand*\circled[1]{\scalebox{0.8}{\tikz[baseline=(char.base)]{
\node[anchor=text, shape=circle,fill, inner sep=0pt, minimum size=1.2em] (char) {\footnotesize \textcolor{white}{#1}};}}}

\newcommand{\distance}{5pt}

%% file: intro.tex
\section{Introduction}

The increasing complexity of source code poses a key challenge to the reliability of large-scale software systems. Software bugs in these systems can lead to safety issues~\cite{bug_safety} for users around the world as well as cause non-negligible financial losses~\cite{bug_loss}. As such, developers have to spend a large amount of time and effort on bug fixing. Consequently, \aprfull (\apr), designed to automatically generate patches to fix software bugs, has attracted wide attention from both academia and industry~\cite{long2016prophet, legoues2012genprog, long2015spr, lou2020can, tufano2018empstudy}.

To achieve \apr, one popular approach is known as Generate-and-Validate (G\&V)~\cite{qi2015gv, ghanbari2019prapr, lou2020can, le2016hdrepair, legoues2012genprog, wen2018capgen, hua2018sketchfix, martinez2016astor, koyuncu2020fixminder, liu2019tbar, liu2019avatar}, which is typically based on the following pipeline: First, fault localization techniques~\cite{wong2016fl, abreu2007ochiai, zhang2013injecting, papadakis2015metallaxis, li2019deepfl, li2017transforming} are applied to determine the suspicious locations in programs where bugs are likely to exist. Then, the buggy locations are used by the \apr tools to generate a list of patches that replace buggy lines with correct lines. Afterward, each patch is validated against the original test suite to identify any \emph{plausible patches} (i.e., passing all tests in the test suite). Finally, to determine the \emph{correct patches}, developers examine the list of plausible patches to see if any of them can correctly fix the bug. 

Traditional \apr tools can mainly be categorized into heuristic-based~\cite{legoues2012genprog, le2016hdrepair, wen2018capgen}, constraint-based~\cite{mechtaev2016angelix, le2017s3, demacro2014nopol, long2015spr} and \template~\cite{ghanbari2019prapr, hua2018sketchfix, martinez2016astor, liu2019tbar, liu2019avatar}. Among these traditional tools, \template \apr tools~\cite{ghanbari2019prapr, liu2019tbar, benton2020effectiveness} have been able to achieve state-of-the-art results. \Template \apr tools typically leverage pre-defined templates (e.g., adding a nullness check) for bug fixing. However, since these fix templates are typically handcrafted, the number and types of bugs they are able to fix can be limited.

To address the limitations of traditional \apr, researchers have proposed various \learning \apr tools~\cite{li2020dlfix, chen2018sequencer, jiang2021cure, lutellier2020coconut, zhu2021recoder, ye2022rewardrepair} based on the \nmtfull (\nmt) architecture~\cite{sutskever2014mt} where the input is the buggy code snippets and the goal is to translate the buggy code snippets into a fixed version. To accomplish this, \learning \apr tools require supervised training datasets with pairs of both buggy and fixed code snippets in order to learn how to perform this translation step. These training data are usually obtained by mining historical bug fixes using heuristics/keywords~\cite{dallmeier2007benchmark}, which can be imprecise for identifying bug-fixing commits; even the actual bug-fixing commits can include irrelevant code changes, leading to further pollution in the dataset~\cite{xia2022alpharepair}.
Moreover, it can be hard for such \apr tools to generalize and fix bug types unseen during training.

To better leverage recent advances in \plmfull{s} (\plm{s}), researchers~\cite{xia2022alpharepair, xia2023repairstudy, kolak2022patch, prenner2021codexws} have directly applied \plm{s} to generate patches without bug-fixing datasets. These \llm-based \apr tools work by either directly generating a complete code function~\cite{prenner2021codexws, xia2023repairstudy} or predict/infill the correct code snippet given its surrounding context~\cite{xia2022alpharepair, xia2023repairstudy}. By directly using \llm{s} that are pre-trained on billions of open-source code snippets, \llm-based \apr tools can achieve state-of-the-art performance on many repair datasets~\cite{xia2022alpharepair}.

Traditional \apr tools have long used the insight of the \emph{plastic surgery hypothesis}~\cite{barr2014plastic} where it states that the code ingredients to fix a bug already exist within the same project. Traditional \apr tools have manually designed pattern-~\cite{ghanbari2019prapr, saha2017elixir} or heuristic-based~\cite{jiang2018simfix, legoues2012genprog} approaches to finding and using such relevant code ingredients to generate fixes for bugs. However, the plastic surgery hypothesis has been largely ignored in \llm-based \apr. In fact, \llm provides a unique opportunity to fully automate the plastic surgery hypothesis idea via fine-tuning (learning project-specific information via model updates from the buggy project) and prompting (directly providing relevant code ingredients to the model), and make it directly applicable to different languages (since the \llm{s} are typically multi-lingual).%
Moreover, despite the intensive manual efforts involved, traditional \apr tools still cannot fully leverage project-specific information due to large search space for leveraging/composing existing code ingredients. In contrast, the project-specific information can effectively leveraged by \llm{s} due to their power in code understanding/vectorization, e.g., even partial/imprecise information may still guide \llm{s} in correct patch generation!
 To this end, we ask the question: \emph{How useful is the plastic surgery hypothesis in the era of \plm{s}}?

\mypara{Our Work.} To answer the question, we present \ourtech{\xspace} -- a \llm-based approach that automatically utilizes the plastic surgery hypothesis by systematically combining multiple fine-tuning and prompting strategies for \apr. \ourtech fine-tunes \plm{s} using two novel domain-specific training strategies: \textbf{\epfinetune} -- we fine-tune using the original buggy project by aggressively masking out a high percentage of tokens, which allows \plm to learn project-specific code tokens and programming styles; and \textbf{\rofinetune} -- which only masks out a single continuous code sequence per training sample, allowing the model to get used to the final \csapr task of predicting a single continuous code sequence. Furthermore, we directly leverage the ability for \plm{s} to understand natural language instructions and introduce a novel prompting strategy, \textbf{\idprompting}, which uses information retrieval and static analysis to obtain a list of relevant identifiers for the buggy lines. While such relevant identifiers are critical for fixing some difficult bugs, they may not be seen by the \llm during inference due to limited context window size. Through the use of prompting, we directly tell the model to use these extracted identifiers (relevant code ingredients) to generate the correct code. Finally, to perform repair, we combine all four model variants (including the base model, both fine-tuned models and the base model with prompting) for the final repair.

While our insight of leveraging the plastic surgery hypothesis for \llm-based \apr is generalizable across different types of \plm{s}, to implement \ourtech, we choose a recent \plm{\xspace}, \ctfive~\cite{wang2021codet5}, which is pre-trained on millions of open-source code snippets. \ctfive is an encoder-decoder model trained using \mspfull (\msp) objective where a percentage of tokens are masked out and each continuous masked token sequence is referred to as a masked span. Also, although we only extract relevant identifiers from the current buggy project (since this paper focuses on the plastic surgery hypothesis), our work can be easily extended to obtain other code information (such as relevant statements or functions) from other sources, such as  the massive pre-training corpora~\cite{husain2020codesearchnet} or historical bug-fixing datasets~\cite{jiang2019infer}, which can provide more coding knowledge for \llm{s}. Besides, although we mainly focus on using traditional string comparison algorithms for information retrieval in this paper, these techniques can be easily replaced by other frequency-based retrieval~\cite{robertson2009probabilistic} and neural search (or embedding-based search)~\cite{reimers2019sentence}.
  In summary, this paper makes the following contributions:

\begin{itemize}[noitemsep, leftmargin=*, topsep=0pt]
    \item \textbf{Dimension.} This paper is the first to revisit the important plastic surgery hypothesis in the era of \llm{s}. It opens up a new dimension for \llm-based \apr to incorporate previously neglected information from the buggy project itself to boost \apr performance. Furthermore, it demonstrates the promising future of retrieval-based prompting for modern \llm-based \apr.
    \item \textbf{Implementation.} We implement \ourtech based on the recent \ctfive model. We augment the model using two novel fine-tuning strategies: \epfinetune and \rofinetune, along with a novel prompting strategy based on information retrieval and static analysis: \idprompting. We combine the patches generated by all four models together and perform patch ranking to speed up \apr.%
    \item \textbf{Evaluation Study.} We conduct an extensive evaluation against state-of-the-art \apr tools. On the widely studied \dfj 1.2 and 2.0 datasets~\cite{just2014dfj}, \ourtech is able to achieve the new state-of-the-art results of 89 and 44 correct bug fixes (15 and 8 more than best baseline) respectively.  Furthermore, we perform a broad ablation study to justify our design. \ourtech demonstrates for the first time that the plastic surgery hypothesis can substantially boost \llm-based \apr and advance state-of-the-art \apr, while being fully automated and general. Moreover, even partial/imprecise code ingredients may still effectively guide \llm{s} for \apr!
\end{itemize}

%% file: background.tex
\section{Background}
\subsection{\plmfull}

\begin{figure}
    \includegraphics[width=0.85\linewidth]{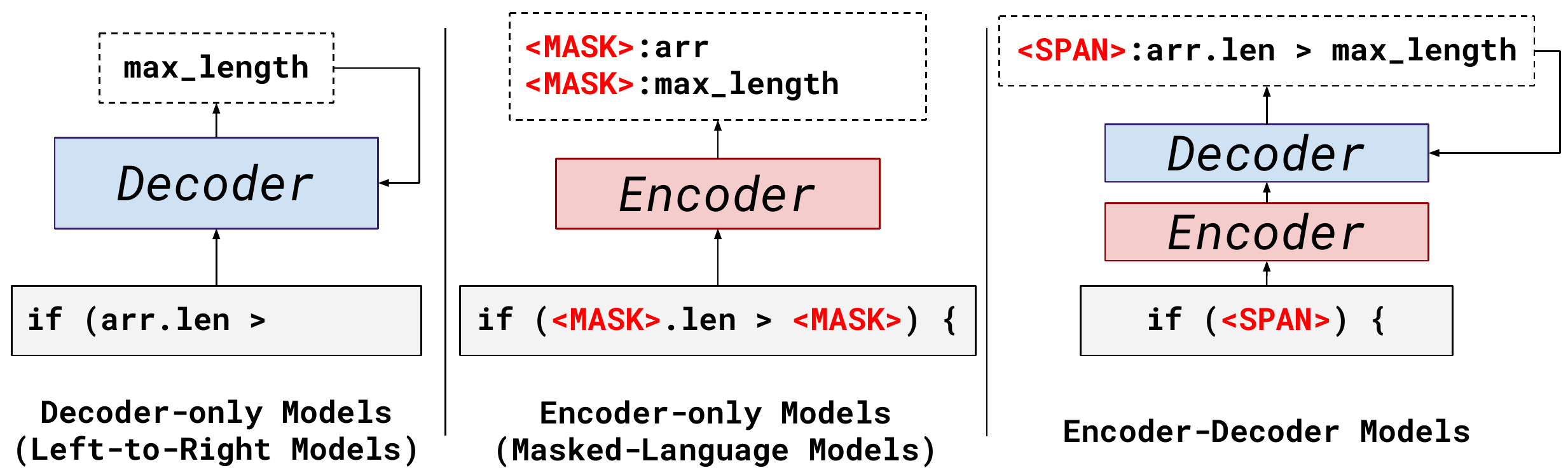}
    \caption{\plm overview}
    \label{fig:plm_overview}
\end{figure}

\plmfull{s} (\plm{s}) trained on large amounts of text combined with code snippets from a broad range of open-source repositories have shown impressive progress in various code-related tasks~\cite{chen2021codex, fried2022incoder, xu2022systematic}. To apply them to code, researchers may further train the models on open-source code repositories. On top of that, \plm{s} can be \emph{fine-tuned} (updating pre-trained model parameters by further training on the downstream tasks~\cite{devlin2018bert}) to target a specific task such as defect prediction~\cite{omri2020defectpred} or code clone detection~\cite{wang2022bridge}. Different from fine-tuning, \emph{prompting} is a way to directly use \llm{s} on downstream tasks without further training. Prompting involves providing a natural language description of the task and optionally including a few demonstrations of the task to the \llm{s} before the actual input. Researchers have successfully applied prompting to tasks like code completion~\cite{Nijkamp2022CG} and code summarization~\cite{kuznia2022summ}.

\plm{s} are based on the popular Transformer architecture~\cite{vaswani2017attention}, which combines an \textbf{encoder} with a \textbf{decoder} to perform text generation. The encoder first takes in the input to the model and then produces an encoded representation. The decoder uses this encoded vector to autoregressively generate the next token based on all previously generated tokens. Using this paradigm, researchers build larger and larger models (as large as 540B in the number of model parameters~\cite{chowdhery2022palm}) and demonstrated impressive results on code-related tasks such as code completion~\cite{chen2021codex} and code synthesis~\cite{austin2021sythesis}. 

\plm{s} can be classified into three groups based on their model architecture and pre-training objective: \textbf{Decoder-only} (Left-to-Right Language Models), \textbf{Encoder-only} (Masked Language Models), and \textbf{Encoder-decoder models}. Figure~\ref{fig:plm_overview} shows an overview of the three different \plm architectures. Decoder-only models perform left-to-right generation by producing the probability of a token given all previous tokens. One of the most well-known \plm{s}, GPT~\cite{radford2019gpt2, brown2020gpt3}, is based on this architecture. During training, decoder-only models aim to predict the next token given all previous context. Examples of decoder-only models for code are CodeGPT~\cite{lu2021codexglue}, CodeParrot~\cite{tunstall2022natural}, and Codex~\cite{chen2021codex}. These models can be directly used for program generation given previous code contexts. Encoder-only models, on the other hand, only use the encoder component to provide an encoded representation of the input. Models such as BERT~\cite{devlin2018bert} are trained using the \mlmfull (\mlm) objective, where a small percentage (e.g., 15\%) of the training tokens are masked out and the model aims to recover these masked tokens using the bi-directional context. CodeBERT~\cite{feng2020codebert}, GraphCodeBERT~\cite{guo2021graphcodebert}, and CuBERT~\cite{kanade2020cubert} are examples of encoder-only models where it can provide a representation of the input code to be used for downstream tasks such as code clone detection~\cite{wang2022bridge}. Encoder-decoder models (T5~\cite{raffel2020t5}, BART~\cite{lewis2019bart}) use both components of the transformer and are typically trained using \mspfull (\msp) objective. Different from \mlm, instead of masking out individual tokens, \msp replaces a sequence of tokens with a single span mask. The goal of the training is to recover the original sequence using both the context before and after the span mask token. For code tasks, \ctfive~\cite{wang2021codet5} and PLBART~\cite{ahmad2021PLBART} are examples of encoder-decoder models and due to the \msp pre-training objective, they can be directly used to fill in arbitrary code snippets given the bi-directional code context.

\subsection{Automated Program Repair}

\aprfull (\apr) works by automatically generating patches when given the buggy project and potential fault locations. Traditional \apr tools can be categorized into constraint-based~\cite{mechtaev2016angelix, le2017s3, demacro2014nopol, long2015spr}, heuristic-based~\cite{legoues2012genprog, le2016hdrepair, wen2018capgen}, and \template~\cite{ghanbari2019prapr, hua2018sketchfix, martinez2016astor, liu2019tbar, liu2019avatar} tools. Among those, \template \apr has been regarded as the state-of-the-art in achieving the best repair performance~\cite{ghanbari2019prapr, liu2019tbar, benton2020effectiveness}. \Template \apr works by using pre-defined templates (handcrafted by human experts) which target specific patterns in source code. Each template will have an associated fix that modifies the found patterns in the source code to fix specific types of bugs. However, \template \apr tools cannot fix bugs that do not fall under the pre-defined templates. As a result, \template tools lack the ability to generalize to unseen bug types.

In recent years, researchers have begun to focus on \learning \apr approaches such as \selfapr~\cite{ye2022selfapr}, \rewardrepair~\cite{ye2022rewardrepair}, \recoder~\cite{zhu2021recoder}, \cure~\cite{jiang2021cure}, and \coconut~\cite{lutellier2020coconut} based on the \nmtfull (\nmt)~\cite{sutskever2014mt} architecture. The goal of these tools is to learn a transformation using DL models that turns buggy code snippets into patched ones. To facilitate this, these tools require further training on specific bug-fixing datasets containing pairs of buggy and fixed code snippets. However, these bug-fixing datasets are usually scraped from open-source bug-fixing commits using handwritten heuristics such as keyword searching~\cite{zhu2021recoder, lutellier2020coconut, dallmeier2007benchmark, jiang2019infer}, which can include irrelevant code commits; even the correctly identified bug-fixing commits may contain various irrelevant code changes (such as refactoring or new feature implementation), introducing various noises in the datasets. Also, to avoid including bug-fixing commits with irrelevant code changes,
existing \learning \apr techniques will limit the commits
to ones with few lines of changes~\cite{jiang2021cure, zhu2021recoder, lutellier2020coconut}, further limiting
the amount of training data. Moreover, \nmt-based \apr tools may still not generalize to specific code or bug types that are not seen inside of the (limited) bug-fixing datasets. 

Recognizing these limitations in \nmt-based \apr, researchers have proposed \llm-based \apr tools~\cite{xia2022alpharepair, xia2023repairstudy, prenner2021codexws, kolak2022patch} which do not require bug-fixing datasets by directly using \llm{s} for \apr. \alpharepair~\cite{xia2022alpharepair} first introduced the \csapr (or infilling-style \apr) to directly leverage \plm{s} in a zero-shot manner to fill in the code given the context before and after the buggy line. \alpharepair first generates repair inputs that replace the original buggy line with masked tokens and uses the \codebert model~\cite{feng2020codebert} to directly recover the correct code to fill in each masked token. Later studies~\cite{xia2023repairstudy, kolak2022patch, prenner2021codexws} also used different \llm{s} (including Decoder-only and Encoder-decoder models) to not only perform \csapr but also repair scenarios where a complete fixed function is generated. Contrasting with \nmt-based \apr tools, \llm-based \apr leverages the pre-training objectives of \plm{s} which can directly learn the relationship between correct code and its context without relying on historical bug-fixing commits. As a result, \llm-based \apr tools have shown to achieve state-of-the-art performance on repair tasks across multiple programming languages~\cite{xia2023repairstudy}. In this work, we present the first work to further advance state-of-the-art \llm-based \apr with the insight of the plastic surgery hypothesis.

%% file: approach.tex
\section{Approach}

\begin{figure}
    \includegraphics[width=0.9\linewidth]{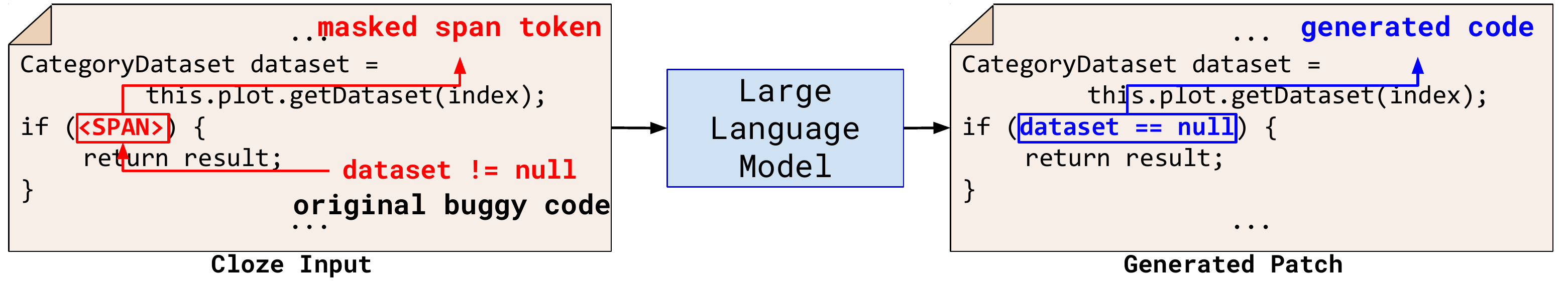}
    \caption{\Csapr}
    \label{fig:csapr}
\end{figure}

\begin{figure*}
    \includegraphics[width=0.8\linewidth]{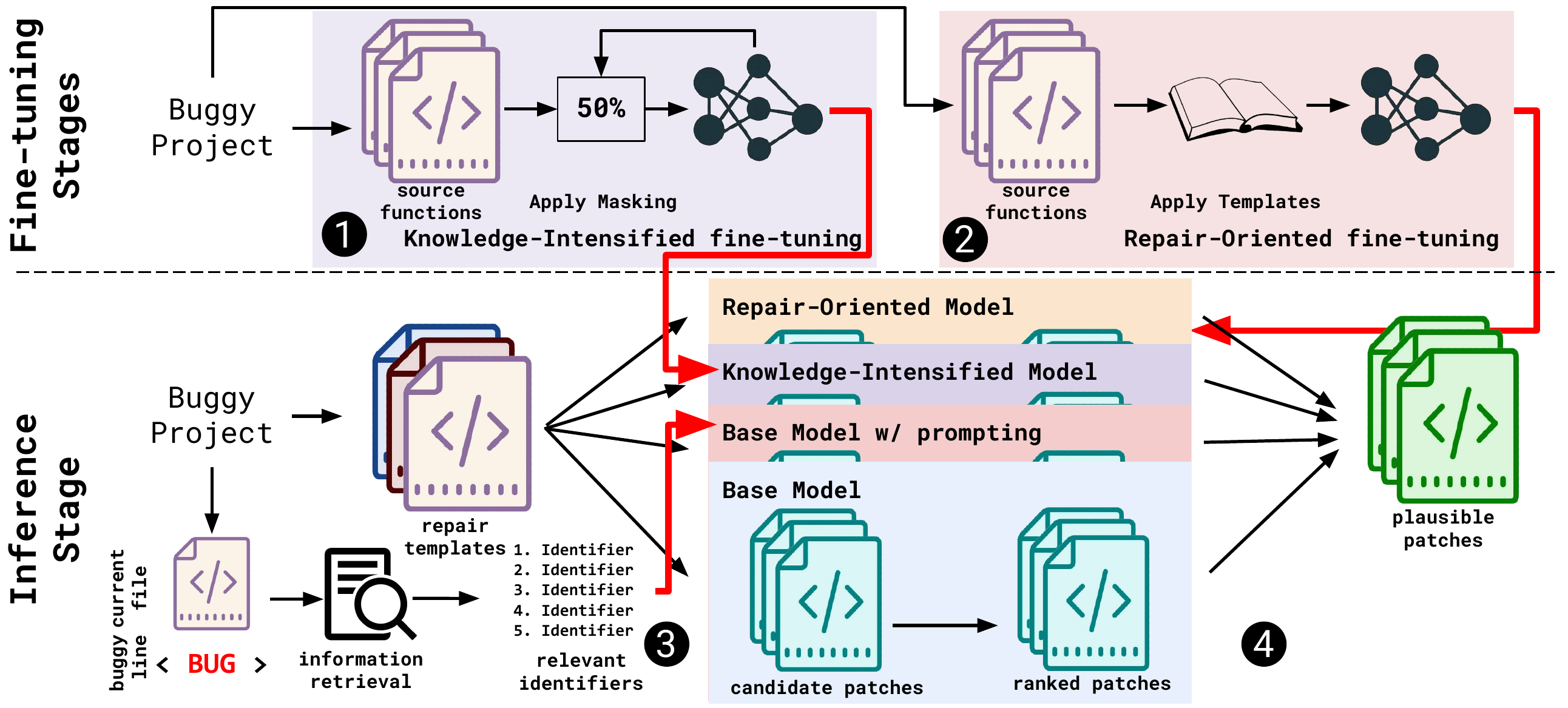}
    \caption{\ourtech overview}
    \label{fig:overview}
\end{figure*}

In this section, we describe our approach to incorporate the plastic surgery hypothesis for \llm-based \apr. While our overall idea is general for \llm-based \apr approaches, we mainly focus on \csapr since it has been demonstrated to be the state-of-the-art for \llm-based \apr~\cite{xia2023repairstudy}. Figure~\ref{fig:csapr} shows the overview of \csapr where we aim to directly generate the correct code in place given the original context, where the model has to "fill in the blanks" of the missing code line/hunk (represented by \CodeIn{<SPAN>}) given the buggy context. However, prior \csapr tools largely ignore the rich project-specific and bug-specific information which has been demonstrated by the plastic surgery hypothesis~\cite{barr2014plastic, legoues2012genprog} to be critical in helping to generate a correct fix. \llm presents an exciting opportunity to automatically leverage the idea of the plastic surgery hypothesis via its powerful ability to both directly learn from the buggy project (fine-tuning) and use relevant code context clues (prompting) to generate correct fixes. In this work, we explore using \llm{s} to automatically capture project-specific information via both fine-tuning and prompting.

We propose a novel approach -- \ourtech to combine the direct usage of \plm in \csapr with knowledge gained from the plastic surgery hypothesis.
\ourtech first trains two separate models with two novel fine-tuning strategies:
\textbf{1) \epfinetune} -- we use the source files of the original buggy project to construct a similar dataset to pre-training by aggressively masking out large portions (50\%) of the code tokens, which allows \plm to learn project-specific code tokens and programming styles; and%
\textbf{2) \rofinetune} -- we fine-tune another model using the original buggy project to construct a more repair-oriented dataset by masking out only a single continuous code sequence per training sample,
which makes the fine-tuned model become more prepared for the repair task where only a single continuous code sequence needs to be generated. 
Additionally, we propose a novel prompting strategy, \textbf{3) \idprompting}, by obtaining a list of relevant/rare identifiers that are not seen by the model in its immediate context using information retrieval and static analysis.%

While our approach can be extended to different \plm{s}, in this paper, we use \ctfive~\cite{wang2021codet5}, an encoder-decoder \plm for code trained using \mspfull (\msp) objective. \msp replaces continuous tokens with a single masked span token (\CodeIn{<SPAN>}) and the pre-training task is to recover masked-out code sequences given the surrounding context. Given a sequence of tokens $X = \{x_1, ..., x_n\}$, random sequences of tokens are replaced with a masked span token to produce $X_{masked} = \{x_1, ..., \CodeIn{<SPAN>}, x_n\}$. Let $M = \{m_1, ..., m_k\}$ be the tokens masked out, $M_{<g} = \{m_{1}, m_{2}, ..., m_{g-1}\}$ be token sequence predicted so far where $g \leq k$, $P$ be the predictor (model) which outputs the probability of a token. The \msp loss function can be described as:

\begin{equation}
\mathcal{L}_{\msp} = -\frac{1}{k}\sum_{i=1}^{k} log\;(P\;(m_i\;|\;X_{masked},\;M_{<i})
\end{equation}

To employ \ctfive for \csapr using \msp, we follow previous work~\cite{xia2022alpharepair} and use repair templates to generate mask lines where we replace the entire or parts of the buggy line with a single masked span token. We then use \ctfive to generate the correct code to replace the masked span and create a patch for the bug. Figure~\ref{fig:overview} shows the overview of our approach:

\begin{itemize}[noitemsep, leftmargin=*, topsep=0pt]
	\item \textbf{\circled{1} (Section~\ref{sec:ki_finetune})}: We use \epfinetune to build a training dataset by extracting functions from buggy project source code. We fine-tune the original \ctfive model by first using a high masking rate to aggressively learn project-specific tokens. 
	\item \textbf{\circled{2} (Section~\ref{sec:ro_finetune})}: We use \rofinetune  strategy to fine-tune another model by constructing another training dataset from the buggy project where only a single or partial code line is masked out based on the repair templates. We train the model until convergence and obtain the \rofinetune model. 
        \item \textbf{\circled{3} (Section~\ref{sec:idprompting})}: We use \idprompting strategy to extract relevant identifiers via information retrieval and static analysis. We then create individual prompts with instructions for the model to use the extracted identifiers during patch generation. 
	\item \textbf{\circled{4} (Section~\ref{sec:patch_gen})}: We perform \csapr by using the repair templates from previous work~\cite{xia2022alpharepair} and generate patches by separately using the 4 models (original \ctfive, \epfinetune model, \rofinetune model, and original \ctfive with prompting). Each patch generated is then validated against the test suite to find the list of plausible patches.

\end{itemize}

\subsection{Knowledge-Intensified Fine-tuning}\label{sec:ki_finetune}

\begin{figure}
    \includegraphics[width=\linewidth]{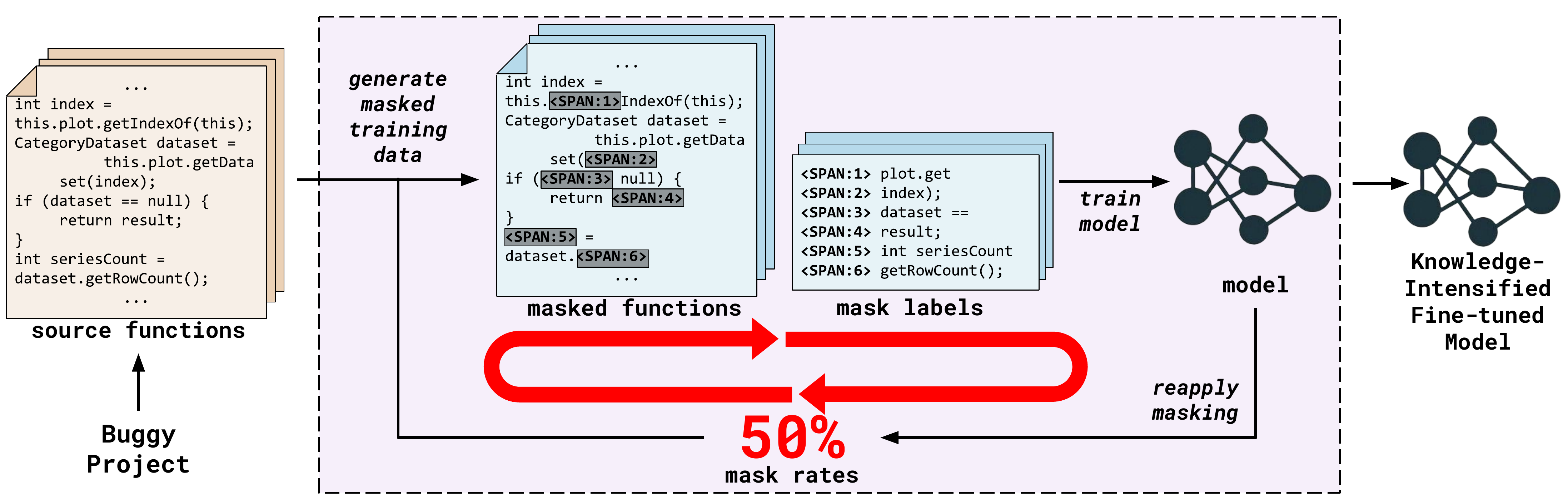}
    \caption{\epfinetune overview}
    \label{fig:ki_overview}
\end{figure}

To facilitate the learning of project-specific information, we use \epfinetune by constructing a training dataset using the buggy project itself. %
Figure~\ref{fig:ki_overview} shows the  \epfinetune process. We first extract the source code functions from the project code base where the bug is from and apply \msp objective -- masking out multiple spans of code tokens, used to pre-train the original \ctfive model. Traditionally, \msp objective will mask out only a small portion of the original code tokens (e.g., 15\%~\cite{wang2021codet5}). However, in this step, we employ a much higher masking rate (50\%) which means the model is tasked with recovering more masked-out code tokens with less context. This approach is motivated by a recent study~\cite{wettig2022mask15} on \plm{s} for natural languages where better representation and downstream performance can be achieved by increasing the pre-training masking rate of \mlm and \msp objectives. The study found that higher masking rates make the learning tasks more challenging and can force the model to learn more aggressively, which helps improve the performance of \plm{s} on various downstream tasks. In this paper, we leverage this idea of more aggressive training to force the model to learn more project-specific knowledge by trying to recover more code tokens given limited context.

However, one limitation with fine-tuning the model on the buggy project itself is the relatively small number of training samples (e.g., thousands of functions) especially compared with the large amount of open-source pre-training data used in \ctfive (millions of functions). As a result, we \textit{reapply} the \msp objective across iterations to augment the fine-tuning dataset with more training samples. Following the example in Figure~\ref{fig:ki_overview}, we start by creating one set of training data by masking out 50\% of the training tokens to create masked spans. In the next iteration, we reapply this masking strategy to create a new set of training data by randomly choosing another 50\% to mask out again. In this process, we essentially create new training data for each subsequent iteration. While the number of tokens masked out is the same, the specific masked locations can be different which provides further augmentation on the training dataset allowing the model to learn more project-specific tokens.

During the fine-tuning process when using \epfinetune, \ourtech is able to learn project-specific knowledge such as commonly used methods or variables that are specifically defined in the current project. These pieces of project-specific knowledge are especially important for repair as many bugs can be fixed by applying code snippets found in other parts of the source file or project, according to the plastic surgery hypothesis~\cite{barr2014plastic}. Due to the limited context window size (e.g., 512 tokens for \ctfive), \ctfive cannot encode all of the surrounding contexts during inferencing, which leads to the base \ctfive model missing variable names and method calls used in other parts of the context that are actually necessary to be used as part of the patch. \epfinetune can partially alleviate this by learning these missing variable names and method calls as part of the fine-tuning such that when used for \csapr, the fine-tuned model can predict these useful tokens with a higher probability compared to the model without fine-tuning. While \epfinetune can help with better learning of project-specific details, we next introduce \rofinetune which produces another model that aims to optimize for the repair task.

\subsection{Repair-Oriented Fine-tuning}\label{sec:ro_finetune}

\begin{figure}
    \includegraphics[width=\linewidth]{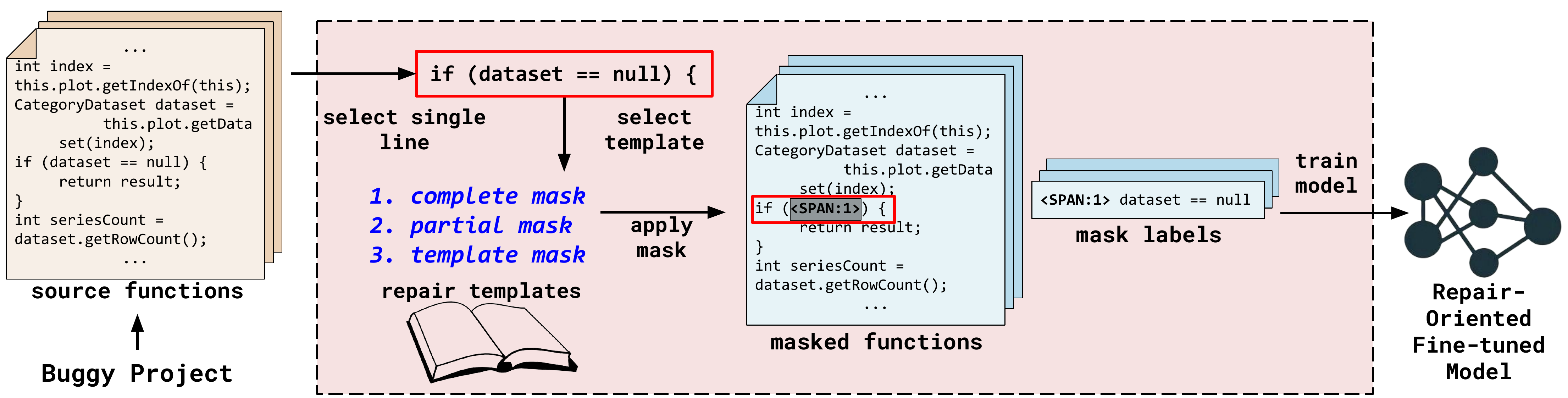}
    \caption{\rofinetune overview}
    \label{fig:ro_overview}
\end{figure}

In the previous step, we use \epfinetune to generate more code that uses project-specific variables, method calls, and structures. \Csapr frames the problem of patch generation by asking the model to fill in the correct code given the buggy context. This is achieved by exploiting the similarity between the pre-training objective and final inference setup to generate patches. Furthermore, \csapr is not limited to only generating a complete line but additionally using repair templates (e.g. replaces only method call name) which keeps part of the buggy line to create patches by generating partial lines~\cite{xia2022alpharepair}. However, both the \epfinetune \ctfive and the original \ctfive suffer from the same limitation: \textit{the training process is not designed for repair}. Both the original pre-training and \epfinetune use \msp which masks out multiple disjointed code token spans. The goal of the model during training is to recover the original tokens for all the masked spans. However, for \csapr, only a single code line or a part of the code line is usually masked out and the model only has to predict the correct code for that single span.

To address such limitations, we use \rofinetune which fine-tunes with a training setup that is similar to the repair inference task. Figure~\ref{fig:ro_overview} shows the detailed \rofinetune process. We again first use the original buggy project as the source of our training data by extracting source code functions. We pick a single code line in each training sample to mask out with a single span token. One can think of this chosen code line as the buggy line in the final repair scenario. To model the impact of eventually using template-based repair inputs, we randomly select a repair template that can be applied on this line to mask out only a part of the line. These repair templates are taken directly from previous \csapr work~\cite{xia2022alpharepair} and can be categorized into 3 different types of repair templates: \textbf{1) Complete mask} -- replace the entire buggy line with a single span token or add the span token to before/after the buggy line, \textbf{2) Partial mask} -- keep some original buggy line tokens at the end or beginning of the line and replace the rest with a span token, and \textbf{3) Template mask} -- target specific code line types by replacing the method call, method parameters, and Boolean expression or operator with a span token.%
In this fine-tuning process, our training samples are closely similar in their setup compared with the \csapr task that we want the model to perform. Using \rofinetune, we can produce a fine-tuned model which aims to follow closely with the final repair task. 

\subsection{Relevant-Identifier Prompting}
\label{sec:idprompting}

Two previous fine-tuning strategies aim to fine-tune the model towards generating more project-specific tokens and get used to repair-oriented inputs, in both cases, using the original buggy project as the fine-tuning dataset. However, this means that the two fine-tuned models have been geared toward the entire buggy project rather than the specific bug within the project. For specific bugs, the relevant code ingredients may be drastically different depending on the bug file, location, and type of buggy line. These ingredients can be possibly far away from buggy locations, making it hard for them to be included in the input context due to the \textbf{limited context window size} of \plm{s} (e.g. the limit of context window size of \ctfive is 512 tokens).

We use \idprompting on the base \ctfive model to directly prompt the model with relevant code ingredients. Algorithm \ref{algo:ri_strategy} shows the overview of our prompting strategy. Given the buggy line information, we first extract the certain file containing the bug and separate it into individual code lines (Line \ref{algo:extractcodelines}). Prior work has found that a significant percentage of the correct code to fix the bug can be found within the same file~\cite{barr2014plastic}. We then use Levenshtein Distance Ratio~\cite{levenshtein1966binary} (other string comparison methods~\cite{ratcliff1988pattern, jaro1989advances, winkler1990string} provide similar results) to measure the similarity between each line compared with the buggy line (Line \ref{algo:calsim}). The hypothesis is that useful identifiers can be obtained from lines that are very similar to the buggy line~\cite{asad2019impact}. We rank each code line based on its string similarity score from high to low to get a ranked list of code lines (Line \ref{algo:rerankcodelines}).
Since we want to provide the model with identifiers to help generate the correct fix, we extract identifiers from each line (Line~\ref{algo:extractid}), which provides us with a ranked list of identifiers. After that, we perform further filtering by first removing any common/simple identifiers (e.g., \CodeIn{length} and \CodeIn{node}) (Line \ref{algo:simplefilter}) and then using static analysis (Line \ref{algo:accessible}) to remove any identifiers that are unaccessible within the buggy method (Line \ref{algo:remove}). Next, we extract the useful type information for each identifier to indicate the type/return type and if it is a method invocation or a variable (Line \ref{algo:typeinfo}). Finally, we obtain a ranked list of complex identifiers that come from similar lines within the same file. 

Using the ranked identifier list, we then generate the prompts to instruct the model to use these extracted identifiers to generate patches (Line \ref{algo:prompt}). \llm{s} are able to understand natural language instructions in the form of prompts to perform specific tasks. In \idprompting, we construct a prompt in the form of \CodeIn{/* use \{\} in the next line */} where we replace \CodeIn{\{\}} with an identifier with type information (e.g., \CodeIn{(Plot) getParent()}). This prompt is then appended before the masked span token during inference which allows the model to directly use this identifier information provided in its generation. Since we have a ranked list of identifiers, we generate multiple unique prompts, each including one of the top \(N\) highest-ranked identifiers. By directly providing these extracted bug-specific identifiers in prompts, the model can use these identifiers which previously are outside of its immediate context to generate the correct fix. 

\begin{algorithm}[t]
    \caption{Relevant-Identifier Prompting Strategy}
    \label{algo:ri_strategy}
    \begin{algorithmic}[1]
    {\small{
    \item[\textbf{Inputs:}] Buggy project $Proj$, Buggy file $File$, Buggy line $buggy\_line$.
    \item[\textbf{Output:}] Relevant-Identifier $prompts$.
    
    \STATE $lines$ $:=$ \textsc{ExtractLines}\ ($File$)  \label{algo:extractcodelines}
    \STATE $similarities,\; identifiers$ $:=$ $[], []$ 
    \FOR{\ $line$\ \  in\ \  $lines$\ }
        \STATE $similarities$.append\ (\textsc{LevenshteinRatio}\ ($buggy\_line$,\; $line$)) \label{algo:calsim}
    \ENDFOR
    \STATE $lines\_ranked$ $:=$ \textsc{RankLines}\ ($lines$,\; $similarities$) \label{algo:rerankcodelines}
    \FOR{\ $line$\ \  in\ \  $lines\_ranked$\ }
        \STATE $line\_identifiers :=$ \textsc{ExtractIds}\ ($line$) \label{algo:extractid}
        \STATE $identifiers$.extend\ (\textsc{SimpleFilter}\ ($line\_identifiers$)) \label{algo:simplefilter}
    \ENDFOR
    \STATE $accessibles$ $:=$ \textsc{FindAccessibleIds}\ ($Proj$,\; $File$,\; $buggy\_line$) \label{algo:accessible}
    \STATE $relevants := identifiers\; \cap\; accessibles$ \label{algo:remove}
    \STATE $type\_infos$ $:=$ \textsc{FindTypeInfo}\ ($Proj$,\; $relevants$) \label{algo:typeinfo}
    \STATE $prompts := $ \textsc{BuildPrompts}\ ($relevants$,\; $type\_infos$) \label{algo:prompt}
    }}
    
    \end{algorithmic}
\end{algorithm}

\subsection{Patch Generation, Ranking, and Validation}
\label{sec:patch_gen}

We directly use the base \ctfive model (with and without prompting) and the two fine-tuned models generated from previous steps for patch generation. To generate patches to replace the buggy lines, we apply repair template inputs from previous work~\cite{xia2022alpharepair} and ask each model to fill in the masked-out span token with generated correct code. Following prior work~\cite{prenner2021codexws, xia2023repairstudy, kolak2022patch}, we sample each model in parallel to generate its own set of patches. In total, for each bug, we generate four lists of potential patches using the four models.

The rankings of patches are computed based on the outputs of each model. We follow the same process as previous work~\cite{xia2022alpharepair} and compute the entropy score using the model. For each candidate patch, we want to provide a likelihood score (entropy) that can accurately reveal the extent to which \ctfive will generate this patch. Let $T = \{t_1, t_2, ..., t_n\}$ be the list of tokens generated for a patch and $C(t_i)$ be the probability of generating token $t_i$ according to \ctfive, then the likelihood score is defined as: $score(T)=\frac{1}{n}\sum_{i=1}^{n}\log\;(C(t_i))$.

We compute this likelihood score for all the patches generated across all the templates and re-rank patches from the highest score to the lowest score. We then validate each candidate patch in accordance with the ranking results. Since each model (two fine-tuned models, base \ctfive with and without prompting) generates its own separate list of patches, we also perform the patch validation in parallel. As such, we can reduce the patch validation time by stopping after any one of the models found a correct patch according to manual inspection by developers. For one bug, \ourtech aims to produce a list of plausible patches that pass the entire test suite and a correct patch that correctly fixes the underlying bug. Since developers can stop this validation process whenever they find a patch to be the correct fix, the overall correct patch ranking is the \emph{minimum} ranking of the correct patches in the ranked patch lists produced by the four models.

%% file: methodology.tex
\section{Experimental Design}
\subsection{Research Questions}
In this paper, we study the following research questions:
\begin{itemize}[noitemsep, leftmargin=*, topsep=0pt]
	\item \textbf{RQ1}: How does \ourtech compare against the state-of-the-art \apr tools?
	\item \textbf{RQ2}: What is the impact of different configurations of \ourtech{}?
	\item \textbf{RQ3}: How does \ourtech generalize in fixing additional bugs from different projects? 
\end{itemize}

We first demonstrate the repair effectiveness of \ourtech against state-of-the-art \apr tools on the popular \dfj 1.2~\cite{just2014dfj} dataset. We study not only the number of bugs fixed in total but also the number of unique bugs fixed compared with previous techniques. Furthermore, we analyze the improvement in terms of patch ranking -- to validate correct patches faster when using \ourtech. Next, we conduct an extensive ablation study on the different configurations of both our two fine-tuning strategies and one prompting strategy. Due to the time cost to train multiple models and generate patches, for the ablation study, we focus on the \CodeIn{Closure} project in \dfj 1.2 which is the largest in terms of both the number of bugs and the size of source code base. Finally, following prior work~\cite{xia2022alpharepair, xia2023repairstudy}, we evaluate against the state-of-the-art \apr tools on the \dfj 2.0~\cite{just2014dfj} dataset to illustrate that \ourtech is not simply overfitting to the 1.2 version.

\subsection{Implementation}
\ourtech is implemented in Python using the PyTorch~\cite{PyTorchWebPage} implementation of the \ctfive model from Hugging Face~\cite{HuggingFaceWebPage}. Our fine-tuning method is based on the pre-trained CodeT5-large (770M) checkpoint. We use JavaParser~\cite{javaparser} to perform static analysis of filtering inaccessible identifiers in scope.  For both \epfinetune and \rofinetune, we repeat the fine-tuning process for 10 iterations by default to augment our fine-tuning dataset. We fine-tune the \ctfive model on an NVIDIA RTX A6000 with 48GB memory using FP16. We use the following set of hyper-parameters to train models: 32 batch size and 1e-4 learning rate with 15K training steps. We use Adam optimizer~\cite{ba2014adam} to update the parameters and use a linear learning rate scheduler with 10\% warmup proportion. For both fine-tuning strategies, we extract the oldest version of the buggy project for training and use the fine-tuned models to generate patches for all bugs in that project. For \idprompting, we use the top 5 most relevant identifiers. For repair, we sample each model 5000 times and validate the top 1000 unique patches produced by each model (at most 4000 patches in total per bug) which is comparable to other baselines. To generate more unique patches for each sample, we use nucleus sampling with top \(p\) of 1 and temperature of 1. We validate the patches on a workstation using AMD Ryzen Threadripper PRO 3975WX CPU with 32-Cores and 256 GB RAM, running Ubuntu 20.04.5 LTS. Similar to prior work~\cite{xia2022alpharepair, jiang2021cure, li2020dlfix}, we use an end-to-end time limit of 5 hours to fix one bug. Note that we sum up the time spent on each of our four processes as \ourtech time cost for fair comparison.%

\subsection{Subject Systems}
We use the widely studied benchmark of \dfj~\cite{just2014dfj} -- a collection of open-source bugs found across 15 different projects to evaluate \ourtech. We follow prior work and separate the dataset into \dfj 1.2 and \dfj 2.0. \dfj 1.2 contains 391 bugs (removing 4 depreciated bugs) across 6 different projects and \dfj 2.0 contains 438 bugs across 9 additional projects. While evaluating \ourtech on all 391 bugs in \dfj 1.2, we follow prior work~\cite{xia2022alpharepair} and choose only the 82 single-line bugs in \dfj 2.0 for evaluation (since existing \learning \apr mainly target single-line fixes).

\subsection{Compared Techniques}
We compare \ourtech against 20 different \apr tools including both state-of-the-art \learning and traditional \apr tools. We choose 8 recent \learning \apr tools for comparison: \alpharepair~\cite{xia2022alpharepair}, \selfapr~\cite{ye2022selfapr}, \rewardrepair~\cite{ye2022rewardrepair}, \recoder~\cite{zhu2021recoder}, \cure~\cite{jiang2021cure}, \coconut~\cite{lutellier2020coconut}, \dlfix~\cite{li2020dlfix} and \sequencer~\cite{chen2018sequencer}. \alpharepair is a recently proposed and state-of-the-art \csapr tool that directly uses a pre-trained model with accessible training set (\codebert). Additionally, we also compare against 12 representative traditional \apr tools: \tbar~\cite{liu2019tbar}, \prapr~\cite{ghanbari2019prapr}, \avatar~\cite{liu2019avatar}, \simfix~\cite{jiang2018simfix}, \fixminer~\cite{koyuncu2020fixminder}, \capgen~\cite{wen2018capgen}, \jaid~\cite{chen2017jaid}, \sketchfix~\cite{hua2018sketchfix}, \nopol~\cite{demacro2014nopol}, \jgenprog~\cite{martinez2015automatic}, \jmutrepair~\cite{martinez2016astor}, and \jkali~\cite{martinez2016astor}. Finally, since \ourtech proposes to combine the base \ctfive (with and without prompting) with the two fine-tuned models, we also compare against a baseline where we run the base \ctfive four times with four random different seeds. This is a fair and necessary baseline to compare against as \ctfive can produce different sampling outputs depending on the random seed and a developer who wishes to use our approach of combining the four models together may also allocate the same GPU resource to run \ctfive four times as well. We refer to this baseline in our evaluation as \ctfiveseed.

We evaluate against these baselines on \textit{perfect fault localization} setting, where the ground-truth location of each bug is provided to the repair tool by comparing the reference developer patch with the buggy code. This is the preferred evaluation setting~\cite{lutellier2020coconut, jiang2021cure, zhu2021recoder, tufano2018empstudy} as it eliminates any result differences caused by using different fault localization techniques~\cite{wong2016fl}. We use the standard metrics for \apr comparison of \textit{plausible patches} -- pass the entire test suite and \textit{correct patches} -- semantically equivalent to the reference developer fix. Following all prior \apr work, correct patches are determined by manually inspecting each plausible patch. Also, following common practice in \apr, we directly report the number of correct and plausible bug fix results from previous studies~\cite{xia2022alpharepair, ghanbari2019prapr, liu2019tbar, zhu2021recoder, ye2022rewardrepair}.

%% file: results.tex
\section{Result Analysis}

\subsection{RQ1: Comparison with State-of-the-art}

\begin{table*}[!htbp]
\centering
\caption{Evaluation results of correct fixes on \dfj 1.2}
\label{tab:dfj_result}
\scalebox{0.8}{
\begin{tabular}{lcc|ccccccccc}
    \hline
    \textbf{Project} & \textbf{\ourtech} & \textbf{\ctfiveseed} & \textbf{\alpharepair} &  \textbf{\selfapr} &  \textbf{\rewardrepair} & \textbf{\recoder} & \textbf{\tbar} & \textbf{\cure} & \textbf{\coconut} & \textbf{\prapr} & \textbf{\dlfix} \\
    \hline
    Chart & 8 & 8 & 9 & 7 & 5 & 10 & 11 & 10 & 7 & 7 & 5 \\
    Closure & 29 & 23 & 23 & 19 &  15 & 21 & 16 & 14 & 9 & 12 & 11\\
    Lang & 19 & 18 & 13 & 10 & 7 & 11 & 13 & 9 & 7 & 6 & 8\\
    Math & 24 & 23 & 21 & 22 & 19 & 18 & 22 & 19 & 16 & 10 & 13\\
    Mockito & 6 & 5 & 5 & 3 & 3 & 2 & 3 & 4 & 4 & 3 & 1 \\
    Time & 3 & 3 & 3 & 3 & 1 & 3 & 3 & 1 & 1 & 3 & 2\\
    \hline
    \textbf{Total} & \textbf{89} & 80 & 74 & 64 & 50 & 65 & 68 & 57 & 44 & 41 & 40\\
    \hline
\end{tabular}}
\end{table*}

\subsubsection{Bugs fixed}
\label{sec:bugs_fixed}
\begin{figure}
    \includegraphics[width=0.8\linewidth]{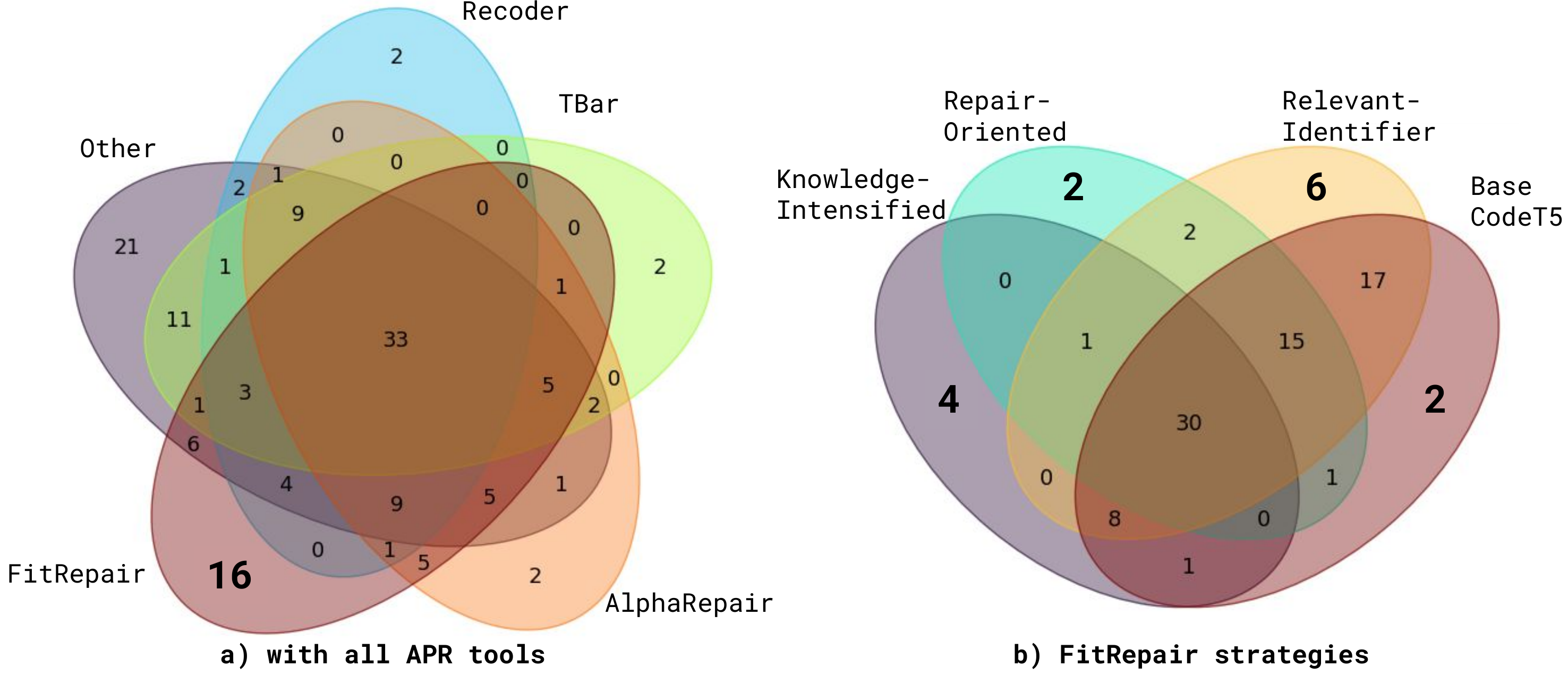}
    \caption{Correct fix Venn diagram on \dfj 1.2}
    \label{fig:venn}
\end{figure}

We first compare \ourtech against both traditional and \learning \apr tools on \dfj 1.2. Table~\ref{tab:dfj_result} shows the number of bugs that can be fixed with correct patches by \ourtech and the top baseline tools. In addition to \ourtech, we also include the result of running the base \ctfive separately four times using four different seeds (Column \ctfiveseed). Compared with \ctfiveseed, we observe that our fine-tuned models and prompting strategy is able to provide additional fixes, boosting the number of correct bug fixes from 80 to 89. In total, \textit{\ourtech is able to achieve 89 correct bug fixes on \dfj 1.2 with 15 more fixes than the current state-of-the-art \apr tool.} Figure~\ref{fig:venn}a shows the number of unique bug fixes (that only one technique can exclusively fix while others cannot) generated by \ourtech compared with the top performing \apr baselines and all other tools (Other). We observe that even compared with all previous \apr approaches, \ourtech is able to \emph{provide 16 additional unique fixes that no other \apr tools have been able to fix so far on \dfj 1.2}.

\begin{figure}
    \includegraphics[width=0.75\linewidth]{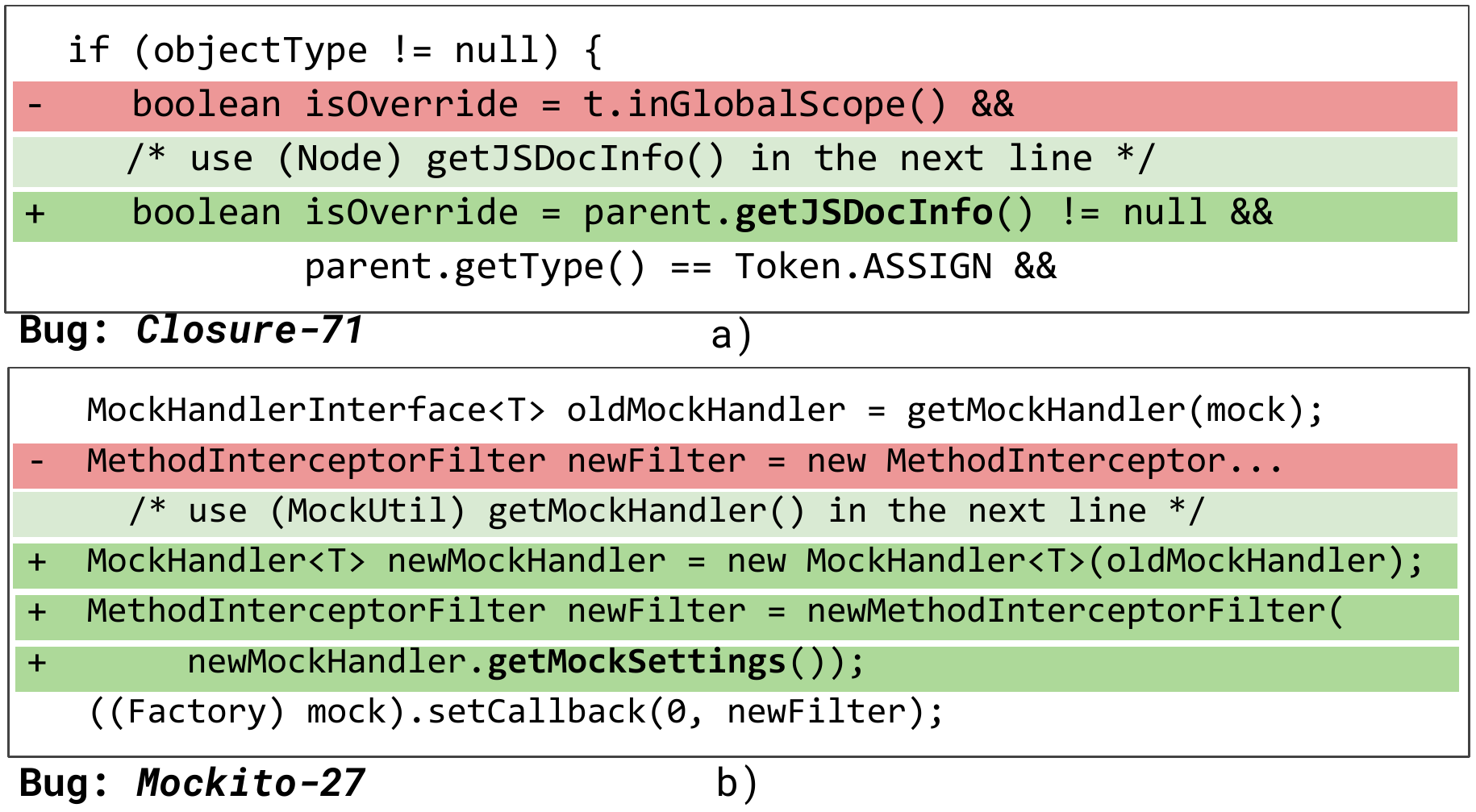}
    \caption{\idprompting unique patches}
    \label{fig:example_unique_fix_idprompting}
\end{figure}

To illustrate the ability of \ourtech, we show an example fix on a bug (\CodeIn{Closure-71}) in Figure~\ref{fig:example_unique_fix_idprompting}a which cannot be fixed by any previous tools. The fix is to invoke the method of \CodeIn{getJSDocInfo()} which is not only an uncommon method name but also is not seen/used in the surrounding context of the bug. However, this method has been used within the same file as the buggy code where another function initializes a variable called \CodeIn{overridingInfo} also using \CodeIn{getJSDocInfo()}. \idprompting is able to recognize the similarity between the buggy variable name (\CodeIn{isOverride}) and this line to extract the relevant identifier of \CodeIn{getJSDocInfo()} and provide the prompt to tell the model to directly use this identifier to generate the correct patch. This example showcases the plastic surgery hypothesis where patches can often be constructed via reusing code snippets/ingredients from other parts of the project. \ourtech directly leverages this hypothesis by extracting the relevant identifiers from similar code lines within the current file and providing it via natural language prompting to generate the correct fix. 

Another very interesting example bug (\CodeIn{Mockito-27}) fixed by \ourtech that cannot be fixed by previous tools is in Figure~\ref{fig:example_unique_fix_idprompting}b. This bug is fixed by calling the \CodeIn{MethodInterceptorFilter} constructor with a previous setting obtained using \CodeIn{getMockSetting()}. We observe that while the \idprompting did not get the exact identifier of \CodeIn{getMockSetting}, it was able to gather a closely related identifier of \CodeIn{getMockHandler}. Due to their power in code understanding/vectorization, \llm{s} do not have to always generate a patch containing the exact identifier in the prompt and can often just use it as hints or partial code for generation. This example further highlights the unique effect the plastic surgery hypothesis have on \llm-based \apr where the extracted code ingredients do not have to be exactly correct and can serve as guidance for the model to generate the correct patch.

\begin{table}[!htbp]
\caption{Average correct patch rank on \dfj 1.2}
\label{tab:patch_ranking}
\scalebox{0.75}{
\begin{tabular}{lcccccc|c}
    \hline
    \textbf{Project} & \textbf{Chart} & \textbf{Closure} & \textbf{Lang} & \textbf{Math} & \textbf{Mockito} & \textbf{Time} & \textbf{Average}\\
    \hline
    \ctfiveseed & 221 & 256 & 100 & 142 & 56 & 342 & 186\\
    \ourtech & 26 & 94 & 76 & 145 & 89 & 260 & 115\\
    \hline
    Improvement & 88\% & 63\% & 24\% & -2\% & -59\% & 24\% & 38\%\\
    \hline
\end{tabular}}
\end{table}

\subsubsection{Individual strategy effect}

\begin{figure}
    \includegraphics[width=0.75\linewidth]{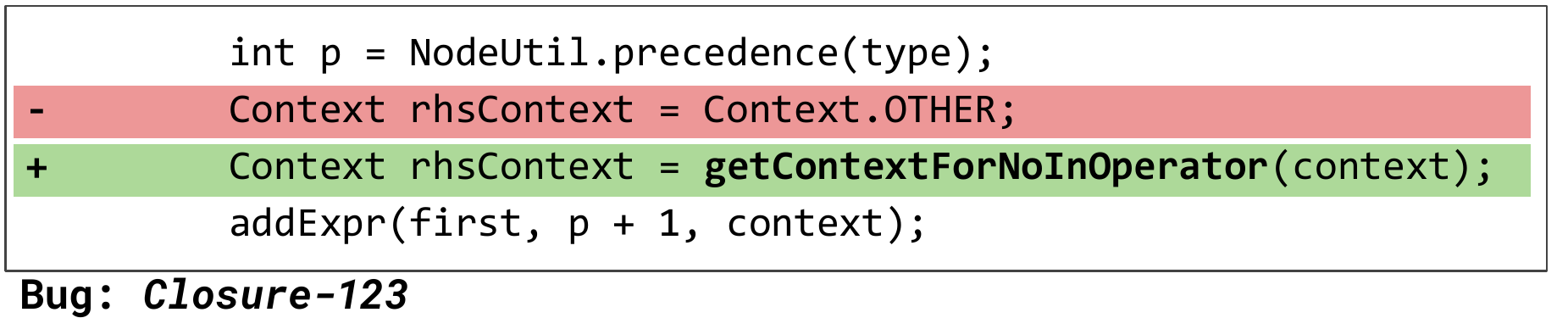}
    \caption{\epfinetune unique patch}
    \label{fig:example_unique_fix}
\end{figure}

Figure~\ref{fig:venn}b shows the unique bugs fixed by each of the individual strategies in \ourtech. We observe that our three strategies are all able to contribute to providing unique fixes compared with the base \ctfive model (4 from \epfinetune, 2 from \rofinetune and 6 from \idprompting) and boost the overall \ourtech to achieve 89 correct patches. Interestingly, each single strategy is already a strong \apr tool, e.g., the \idprompting strategy can fix 79 bugs by itself, already  outperforming all existing tools. This demonstrates the ability of our three strategies to provide additional fixes that directly applying the base \ctfive cannot provide. Moreover, the base \ctfive model without any changes also produced 2 unique fixes, demonstrating the usefulness/necessity of applying the base model to cover the bugs that may only require general-purpose correct code knowledge. 

Since Section~\ref{sec:bugs_fixed} has shown unique fixes obtained by \idprompting, we now present an example bug fixed by \rofinetune. 
Figure~\ref{fig:example_unique_fix} shows a correct fix example that base \ctfive model cannot fix on \CodeIn{Closure-123}. For this bug, the correct fix is to initialize the variable \CodeIn{rhsContext} by calling a function \CodeIn{getContextForNoInOperator()}. What makes this bug difficult to fix for previous \apr tools is that \CodeIn{getContextForNoInOperator} is a very hard sequence to generate. First, it is not a commonly used function name such as \CodeIn{getContext}. Second, there are no code snippets using this function in the immediate context. As such, previous techniques may fail to generate this patch as it requires specific knowledge about the buggy project in order to come up with this function name. However, this function is used multiple times within the buggy project (in other functions and files). \ourtech leverages this by using \epfinetune strategy to fine-tune a model to predict masked-out tokens within the buggy project. During \epfinetune, the model can learn the usage of this specific function within the buggy project and apply it in this case to produce the correct patch. This domain-specific knowledge cannot be learned just from pre-training on a large amount of open-source code (previous \csapr tools).%

\subsubsection{Patch ranking} We examine the ability of \ourtech to perform patch ranking in order to prioritize faster validation for correct patches. Similarly, we compare \ourtech against the baseline of running base \ctfive four times with different seeds.%
Table~\ref{tab:patch_ranking} shows the average rank of the correct patch for the \dfj 1.2 projects on the same set of bugs both \ourtech and \ctfiveseed can fix. We observe that in four out of the six projects, \ourtech provides a better rank on average for the correct patches. On average, using \ourtech, we can achieve a 38\% reduction in the ranking of correct patches. 
\ourtech can learn/use project-specific information to rank correct patches higher since the correct patches often use project-specific identifiers which are less prioritized by the base \ctfive model. In this way, \ourtech not only fixes more bugs, but can also find the correct fixes faster and reduce the computation cost needed for patch validation.

\subsection{RQ2: Detailed Ablation Study}
\label{sec:ablation}

\subsubsection{Impacts of \epfinetune}
\label{sec:ablation_epfinetune}

\begin{table}[htb]
\centering
\caption{Repetition for \epfinetune}
\label{tab:repeat_result}
\scalebox{0.7}{
\begin{tabular}{lcccc}
\hline
\textbf{Strategy} & \makecell{\textbf{\#Corr. / \#Plaus.}\\\textbf{(All)}} & \makecell{\textbf{\#Corr./\#Plaus.}\\\textbf{(New)}} & \makecell{\textbf{Comp.}\\\textbf{Error pct}} & \makecell{\textbf{\#Unique comp.}\\\textbf{per bug}} \\  
\hline
Repetitive (default)       & \textbf{15} / \textbf{30}                                                                                      & \textbf{2} / \textbf{3}                                                                                        & 82\%      & \textbf{138}           \\ 
Non-Repetitive    & 14 / 25                                                                                      & 1 / 2                                                                                        & \textbf{79\%}      & 104            \\ \hline
\end{tabular}}
\end{table}

\begin{table}[htb]
\centering
\caption{Masking rates of \epfinetune}
\label{tab:mask_rate}
\scalebox{0.75}{
\begin{tabular}{lcccc}
\hline
\textbf{Mask Rate} & \makecell{\textbf{\#Corr. / \#Plaus.}\\\textbf{(All)}} & \makecell{\textbf{\#Corr./\#Plaus.}\\\textbf{(New)}} & \makecell{\textbf{Comp.}\\\textbf{Error pct}} & \makecell{\textbf{\#Unique comp.}\\\textbf{per bug}} \\ 
\hline
10\%               & \textbf{16} / 23                                                                      & \textbf{2} / 2                                                                        & 83\%                         & 79            \\ 
20\%               & 14 / 27                                                                      & \textbf{2} / 4                                                                        & 88\%                         & 75            \\ 
30\%               & 14 / 30                                                                      & \textbf{2} / 4                                                                        & 87\%                         & 92            \\ 
40\%               & \textbf{15} / 29                                                                      & \textbf{2} / 4                                                                        & 85\%                         & 102            \\ 
50\% (default)              & \textbf{15} / 30                                                                      & \textbf{2} / 3                                                                        & \textbf{82\%}                & \textbf{138}           \\ 
60\%               & 13 / 26                                                                      & 1 / 4                                                                        & 87\%                         & 106  \\ 
70\%               & 12 / 21                                                                      & 1 / 2                                                                        & 88\%                         & 81            \\ 
80\%               & 9 / 17                                                                       & \textbf{2} / 2                                                                        & 86\%                         & 88            \\ 
90\%               & 7 / 13                                                                       & \textbf{2} / 3                                                                        & 93\%                         & 47            \\ 
\hline
\end{tabular}}
\end{table}

The goal of training using \epfinetune is to incorporate more project-specific knowledge to \ctfive. There are two hyper-parameters of \epfinetune, including \textbf{mask rate} and \textbf{repetition iterations}. Our default setting is to use a 50\% mask rate and 10 repetition times. We conduct an ablation experiment to study the impacts of these two hyper-parameters on the number of correct/plausible patches generated, the number of unique bugs fixed (via correct/plausible patches) when compared to the base \ctfive model, the compilation error rate, and the number of unique compilable patches generated per bug.

We first examine the impact of repeating the masking multiple times during training to generate additional training samples. Table~\ref{tab:repeat_result} shows the results on the Closure bugs with 10 repetition iterations -- generating 10 training sets (Row Repetitive) and no repetition iterations -- generating only 1 training set (Row Non-Repetitive). We observe that the number of total correct and plausible patches produced by the repetitive training approach is higher. Additionally, when we generate new masked training samples during each iteration, the model produced is able to generate two unique bug fixes compared with the base \ctfive model. Similar results can be found when we look at the compilation error rate together with the number of unique compilable patches generated. We see that while the non-repetitive approach has a lower compilation error rate, the number of compilable patches generated is much less. By repeating the masking multiple times during training, we are able to fine-tune the model to learn more project-specific information to produce compilable patches and to fix more unique bugs. %

Next, we study the impacts of different mask rates. In this experiment, we use the default of 10 repetition iterations by generating 10 unique training samples during fine-tuning and examine how different mask rates can have on performance. Due to the extremely large search space (considering unlimited choices of mask rates), we choose mask rates from 10\% to 90\% with an interval of 10\% (9 different mask rates in total). Table~\ref{tab:mask_rate} shows our experimental results on the bugs in the \CodeIn{Closure} project. First, we observe that an extremely high mask rate (70, 80, 90\%) performs poorly in terms of the number of bugs fixed and compilation rate. While a high mask rate may force the model to learn more project-specific tokens during training, each training sample will have a majority of its tokens masked out. Compared with the final repair task of generating a single or partial line, the extremely high mask rate makes the resulting model ill-suited for repair. We observe that the default setting of 50\% mask rate strikes a good balance between achieving the high total number of bugs fixed, more unique bugs fixed compared to base \ctfive, and a relatively low compilation error rate. By using a balanced mask rate of 50\%, the \epfinetune model is able to best complement the base \ctfive in generating more unique bug fixes.

\subsubsection{Impacts of \rofinetune}
\label{sec:ablation_rofinetune}
\begin{table}[htp]
\centering
\caption{Masking strategies of \rofinetune}
\label{tab:strategy}
\scalebox{0.7}{
\begin{tabular}{lccc}
\hline
\textbf{Strategy} & \textbf{AST masking} & \textbf{Single-line masking} & \textbf{Template masking} \\ 
\hline
\#Corr. / \#Plaus. (All) & \textbf{13} / 25      & 10 / 21              & \textbf{12} / 23         \\ 
\#Corr. / \#Plaus. (New) & 0 / 1        & \textbf{1} / 1                & \textbf{1} / 2             \\ 
Comp. Error pct & 88\% & 86\% & \textbf{76\%}\\
\#Unique comp. per bug & 101 & 129 & \textbf{139} \\
\hline
\end{tabular}}
\end{table}

In addition to looking at the impact of different configurations when using \epfinetune, we study the different ways we can apply \rofinetune. Specifically, we design two additional strategies that can be used during training to produce masked training samples. Table~\ref{tab:strategy} shows the result of our default ``Template masking'', ``AST masking'', and ``Single-line masking'' on bugs in the \CodeIn{Closure} project. Recall that our default setting applies repair templates that we use for \csapr directly on the training data to produce masked lines. AST masking will parse the selected line into an AST and randomly choose a subtree to mask out. On the other hand, single-line masking will simply mask out one entire line in a training sample. We observe that single-line masking performed the worst in terms of the number of correct and plausible patches. This is due to the fact that during repair, we use repair templates that not just regenerate complete lines but also mask out part of the lines. The model just has to regenerate the partial code within the line. Single-line masking is only trained on generating the complete line and thus does not perform well when used with repair templates. Additionally, when compared with AST masking, template masking is able to fix more unique bugs compared with the base \ctfive since it directly leverages the inference repair templates to create training samples. While AST masking makes use of the structure information, it does not fully emulate the inference setting of \csapr. Furthermore, the two baselines both resulted in a lot of patches with high compilation error rates compared with our proposed template masking strategy. Template masking is able to directly learn the types of repair templates that the final repair task will use as input for the model, resulting in fewer compilation failures. By using template masking for \rofinetune, we can train a model that is optimized for the repair task in order to generate more bug fixes to compliment the base \ctfive model, which shows that fine-tuning model with training strategies that assemble the underlying repair techniques is able to further boost its bug-fixing performance.%

\subsubsection{Impacts of \idprompting} 

\begin{table}[htp]
\centering
\caption{Configurations of \idprompting}
\label{tab:idprompting_confg}
    \scalebox{0.75}{
    \begin{tabular}{lcc}
    \hline
    \textbf{Configuration} & \makecell{\textbf{\#Corr. / \#Plaus.}\\\textbf{(All)}} & \makecell{\textbf{\#Corr./\#Plaus.}\\\textbf{(New)}} \\ 
    \hline
    \textbf{Default(top-5 current-file type separate)} & \textbf{25 / 39} & \textbf{3 / 3}\\
    \hline 
    Top-1 & 24 / 37 & 1 / 1\\
    Top-10 & 23 / 38 & 1 / 1\\
    Top-20 & 23 / 37 & 1 / 1\\
    \hline 
    Full project & 23 / 36 & 1 / 1\\
    No type & 24 / 37 & 2 / 2\\
    Together & 24 / 40 & 1 / 1\\
    \hline
    \end{tabular}}
\end{table}

We examine the different parameters of our \idprompting strategy. Table~\ref{tab:idprompting_confg} shows the results of our default (Row Default) approach and other configurations on \CodeIn{Closure} project. We first look at the effect of varying the top \(N\) identifiers and observe that when only considering the top-1 identifier for each bug we do not generate more correct fixes since it is unlikely that the relevant identifier to fix the bug always has the highest ranking. On the flip side, considering a larger amount of identifiers per bug (top-10, top-20) is also not desirable since we limit the model to sample only 5000 times per bug, and generating more prompts will decrease the number of samples per each prompt. Next, we look at the scope of the project where we find relevant tokens. Our default setting considers only the current file of the bug and we compare this to when we consider the full project (i.e., changing Line~\ref{algo:extractcodelines} of Algorithm~\ref{algo:ri_strategy} to consider all files within the project). We observe that the number of correct and plausible fixes decreases which reflects a similar finding from prior work~\cite{barr2014plastic} where a significant amount of correct fixing ingredients (relevant identifiers) can already be found within the same file. Furthermore, by considering the entire project, we could introduce more noise where potentially irrelevant identifiers could be highly ranked. Following, we compare the effect of having type information of the identifier in the prompt. We observe that our default setting (with type) is able to generate more correct fixes compared to without types, indicating the usefulness of such information in helping the model generate the correct usage of the identifier in the patch. Finally, we examine our default prompting method of only providing one relevant identifier at a time. We compare this against another approach to include all the top 5 relevant tokens in the same prompt. We see that separating each relevant identifier to its own prompt provides us with more fixes as including all identifiers together can potentially confuse the model. 

\subsubsection{Overhead of \ourtech}

As \ourtech proposes to fine-tune two separate models along with prompting via information retrieval and static analysis, we investigate the extra overhead of using \ourtech compared to just using the base \ctfive. Recall that \ourtech only fine-tunes the model on the oldest version of the project for \dfj 1.2 (one-time cost) and uses the trained models to generate patches for all bugs within that project. We find that on average, for each bug in \CodeIn{Closure}, \ourtech adds 14.3 minutes (6.6 for each fine-tuning strategy and 1.0 for prompting strategy compared with directly using the base \ctfive model. This shows that overall, \ourtech adds a minimal amount of overhead to the repair process (still within the 5-hour limit including overhead). For practical use, the fine-tuning steps can be done ahead of the actual repair task (e.g., periodically during nights or weekends), incurring no additional time cost compared to previous \llm-based \apr tools. Developers can then apply the fine-tuned models together with the base model whenever a bug is detected.

\subsection{RQ3: Generalizability of \ourtech}

\begin{table}[!htbp]
\centering
\caption{Evaluation results of correct fixes on \dfj 2.0}
\label{tab:dfj2_result}
\scalebox{0.8}{
\begin{tabular}{ccccccc}
    \hline
    \makecell{\textbf{\ourtech}} & \textbf{\ctfiveseed} & \makecell{\textbf{Alpha}\\\textbf{Repair}} & \textbf{\selfapr} & \makecell{\textbf{Reward}\\\textbf{Repair}} & \textbf{\recoder} & \textbf{\tbar}\\
    \hline
     44 & 42 & 36 & 31 & 25 & 11 & 8\\
    \hline
\end{tabular}}
\end{table}

\begin{figure}
    \includegraphics[width=0.75\linewidth]{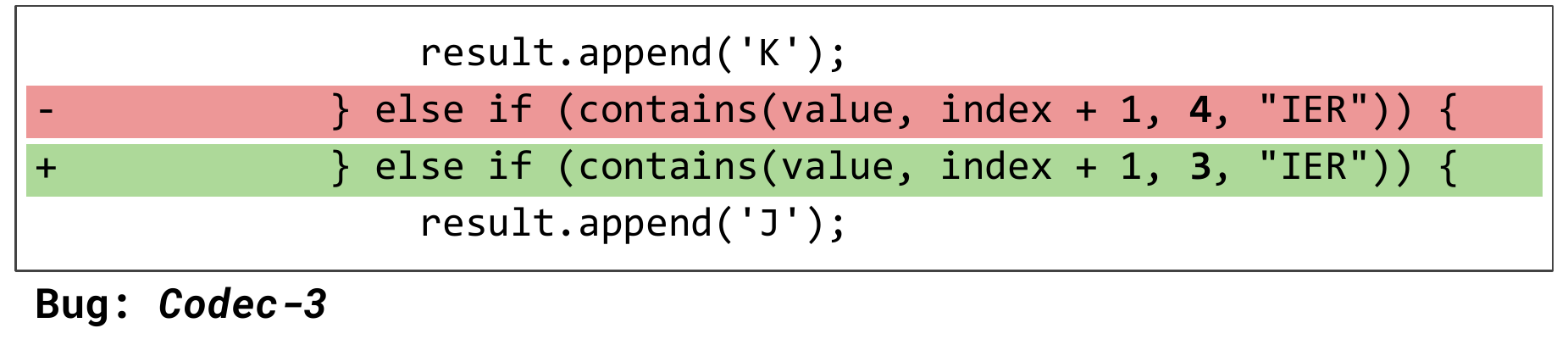}
    \caption{\rofinetune unique patch}
    \label{fig:example_unique_fix_dfj}
\end{figure}

We further evaluate the generalizability of \ourtech on an additional repair dataset of \dfj 2.0 containing new bugs and projects. Table~\ref{tab:dfj2_result} shows the number of correct bug fixes on single-line bugs in \dfj 2.0. We observe that \ourtech is able to achieve the state-of-the-art with the highest number of correctly fixed bugs of 44 (8 more than the best-performing baseline). Unlike other \nmt-based or traditional \template \apr tools, \ourtech does not suffer from the dataset overfitting issue of only performing well on the base \dfj 1.2 dataset. In fact, the relative improvement in the total number of bugs fixed is higher on \dfj 2.0 (22.2\% increase) compared to 1.2 (20.3\% increase). Furthermore, comparing against the baseline (Column \ctfiveseed), \ourtech is able to improve the number of total bug fixes from 42 to 44 and produce 3 unique bug fixes. 

Figure~\ref{fig:example_unique_fix_dfj} shows a bug (\CodeIn{Codec-3}) fixed by \ourtech but cannot be fixed by any other studied \apr tool. The root cause of this bug is an off-by-one error on the second argument in the function call. While this bug looks very simple to fix, one reason previous \learning \apr was not able to provide a correct fix could be the unconventional values of "4" and "3". During the training, \nmt-based \apr can learn from bug-fixing datasets where it is common to use swap a "0" to a "1" and vice-versa to fix a bug. However, changing a "4" to a "3" can be uncommon in the bug-fixing dataset. \Csapr tool that directly leverages \plm{s} can also have a hard time on this bug since the change is very small even if a direct repair template can be applied. Since this change is very small, the \llm should not add any additional code other than changing "4" to "3". However, during training one single mask span usually represents multiple different tokens, which may cause the base \ctfive model to generate more tokens than needed. Using \ourtech and specifically the \rofinetune strategy, the fine-tuned \ctfive can learn such short code generation that usually stems from repair templates such as argument replacement. As such, \ourtech is able to generate this simple patch to fix the underlying bug.%

%% file: threat_to_validity.tex
\section{Threats To Validity}

\mypara{Internal.} Our manual examination in determining the correct patches, semantically equivalent to reference developer patches, from plausible patches is one internal threat to validity. Following common \apr practice, the first two authors perform a careful analysis of each plausible patch along with multiple discussions to determine the correctness of a patch.%

Another internal threat to validity comes from using the \ctfive model which is trained on open-source GitHub code snippets~\cite{husain2020codesearchnet}. This means the training data could overlap with our evaluation repair dataset of \dfj. To address this, we follow prior work~\cite{xia2022alpharepair} and compute the number of patched functions by \ourtech that also exist in the pre-training dataset. In total, out of the 89 bugs fixed on \dfj 1.2, 13 of these fixes are part of the original pre-training dataset of \ctfive. This shows that the majority of the correct fixes (76/89 = 85\%) do not contain any reference developer patch in the training data. Furthermore, for a fair comparison, if we exclude the 13 bugs whose patched functions overlap with the \ctfive pre-training dataset following prior work~\cite{xia2022alpharepair}, we are still able to achieve state-of-the-art performance on \dfj 1.2 with 76 total fixes compared to 67 from the best-performing baseline on the remaining bugs. This shows that \ourtech is not simply performing well on the datasets due to the developer reference patches that the model saw during pre-training. Similarly, on \dfj 2.0, we found that 6 fixed bugs have their reference patch function within the training dataset. Applying the same removal comparison, we still achieve the state-of-the-art result of 38 compared to the best-performing baseline of 30 on \dfj 2.0. Additionally, we also demonstrate that regardless of the overlap between the training and evaluation datasets, by combining project-specific fine-tuning and prompting strategies, we can further improve the performance of the base \llm. Future work to completely address this threat would need to retrain the \ctfive model from scratch after removing the overlapping functions.

\mypara{External.} The major external threat to validity comes from our evaluation dataset. The performance achieved by \ourtech may not generalize well to other datasets. To address this, we use two different versions of \dfj, namely 1.2 and 2.0, and demonstrated that \ourtech is able to achieve state-of-the-art results on both datasets. In the future, we plan on continuing to address this threat by performing more evaluations on other repair datasets across multiple programming languages. 

%% file: conclusion.tex
\section{Conclusion}

In this paper, we have proposed \ourtech, the first fully automated approach to incorporate domain-specific knowledge with the insights of the \emph{plastic surgery hypothesis} for boosting the performance of \plm{s} for APR. \ourtech opens up a new dimension for \llm-based \apr by using both fine-tuning and prompting to combine the power of \llm with project-specific information.
Our evaluation results on the popular \dfj 1.2 and 2.0 datasets show that \ourtech is able to achieve the new state-of-the-art results in fixing 89 and 44 bugs (15 and 8 more than the best-performing baselines) respectively.%
Different from prior \apr work on plastic surgery hypothesis, \ourtech is fully automated, effective, and general. Moreover, even partial/imprecise information may still effectively guide \llm{s} for \apr!

%% file: main.bbl

\begin{thebibliography}{81}


\ifx \showCODEN    \undefined \def \showCODEN     #1{\unskip}     \fi
\ifx \showDOI      \undefined \def \showDOI       #1{#1}\fi
\ifx \showISBNx    \undefined \def \showISBNx     #1{\unskip}     \fi
\ifx \showISBNxiii \undefined \def \showISBNxiii  #1{\unskip}     \fi
\ifx \showISSN     \undefined \def \showISSN      #1{\unskip}     \fi
\ifx \showLCCN     \undefined \def \showLCCN      #1{\unskip}     \fi
\ifx \shownote     \undefined \def \shownote      #1{#1}          \fi
\ifx \showarticletitle \undefined \def \showarticletitle #1{#1}   \fi
\ifx \showURL      \undefined \def \showURL       {\relax}        \fi
\providecommand\bibfield[2]{#2}
\providecommand\bibinfo[2]{#2}
\providecommand\natexlab[1]{#1}
\providecommand\showeprint[2][]{arXiv:#2}

\bibitem[Abreu et~al\mbox{.}(2007)]%
        {abreu2007ochiai}
\bibfield{author}{\bibinfo{person}{Rui Abreu}, \bibinfo{person}{Peter
  Zoeteweij}, {and} \bibinfo{person}{Arjan~J.C. van Gemund}.}
  \bibinfo{year}{2007}\natexlab{}.
\newblock \showarticletitle{On the Accuracy of Spectrum-based Fault
  Localization}. In \bibinfo{booktitle}{\emph{Testing: Academic and Industrial
  Conference Practice and Research Techniques - MUTATION (TAICPART-MUTATION
  2007)}}. \bibinfo{pages}{89--98}.
\newblock


\bibitem[Ahmad et~al\mbox{.}(2021)]%
        {ahmad2021PLBART}
\bibfield{author}{\bibinfo{person}{Wasi~Uddin Ahmad}, \bibinfo{person}{Saikat
  Chakraborty}, \bibinfo{person}{Baishakhi Ray}, {and} \bibinfo{person}{Kai-Wei
  Chang}.} \bibinfo{year}{2021}\natexlab{}.
\newblock \bibinfo{title}{Unified Pre-training for Program Understanding and
  Generation}.
\newblock
\newblock
\showeprint[arxiv]{2103.06333}~[cs.CL]


\bibitem[Asad et~al\mbox{.}(2019)]%
        {asad2019impact}
\bibfield{author}{\bibinfo{person}{Moumita Asad}, \bibinfo{person}{Kishan~Kumar
  Ganguly}, {and} \bibinfo{person}{Kazi Sakib}.}
  \bibinfo{year}{2019}\natexlab{}.
\newblock \showarticletitle{Impact analysis of syntactic and semantic
  similarities on patch prioritization in automated program repair}. In
  \bibinfo{booktitle}{\emph{2019 IEEE International Conference on Software
  Maintenance and Evolution (ICSME)}}. \bibinfo{pages}{328--332}.
\newblock


\bibitem[Austin et~al\mbox{.}(2021)]%
        {austin2021sythesis}
\bibfield{author}{\bibinfo{person}{Jacob Austin}, \bibinfo{person}{Augustus
  Odena}, \bibinfo{person}{Maxwell Nye}, \bibinfo{person}{Maarten Bosma},
  \bibinfo{person}{Henryk Michalewski}, \bibinfo{person}{David Dohan},
  \bibinfo{person}{Ellen Jiang}, \bibinfo{person}{Carrie Cai},
  \bibinfo{person}{Michael Terry}, \bibinfo{person}{Quoc Le}, {and}
  \bibinfo{person}{Charles Sutton}.} \bibinfo{year}{2021}\natexlab{}.
\newblock \bibinfo{title}{Program Synthesis with Large Language Models}.
\newblock
\newblock
\urldef\tempurl%
\url{https://arxiv.org/abs/2108.07732}
\showURL{%
\tempurl}


\bibitem[Barr et~al\mbox{.}(2014)]%
        {barr2014plastic}
\bibfield{author}{\bibinfo{person}{Earl~T. Barr}, \bibinfo{person}{Yuriy Brun},
  \bibinfo{person}{Premkumar Devanbu}, \bibinfo{person}{Mark Harman}, {and}
  \bibinfo{person}{Federica Sarro}.} \bibinfo{year}{2014}\natexlab{}.
\newblock \showarticletitle{The Plastic Surgery Hypothesis}. In
  \bibinfo{booktitle}{\emph{Proceedings of the 22nd ACM SIGSOFT International
  Symposium on Foundations of Software Engineering}} (Hong Kong, China)
  \emph{(\bibinfo{series}{FSE 2014})}. \bibinfo{publisher}{Association for
  Computing Machinery}, \bibinfo{address}{New York, NY, USA},
  \bibinfo{pages}{306–317}.
\newblock


\bibitem[Benton et~al\mbox{.}(2020)]%
        {benton2020effectiveness}
\bibfield{author}{\bibinfo{person}{Samuel Benton}, \bibinfo{person}{Xia Li},
  \bibinfo{person}{Yiling Lou}, {and} \bibinfo{person}{Lingming Zhang}.}
  \bibinfo{year}{2020}\natexlab{}.
\newblock \showarticletitle{On the Effectiveness of Unified Debugging: An
  Extensive Study on 16 Program Repair Systems}. In
  \bibinfo{booktitle}{\emph{ASE}}. \bibinfo{pages}{907--918}.
\newblock


\bibitem[Brown et~al\mbox{.}(2020)]%
        {brown2020gpt3}
\bibfield{author}{\bibinfo{person}{Tom~B. Brown}, \bibinfo{person}{Benjamin
  Mann}, \bibinfo{person}{Nick Ryder}, \bibinfo{person}{Melanie Subbiah},
  \bibinfo{person}{Jared Kaplan}, \bibinfo{person}{Prafulla Dhariwal},
  \bibinfo{person}{Arvind Neelakantan}, \bibinfo{person}{Pranav Shyam},
  \bibinfo{person}{Girish Sastry}, \bibinfo{person}{Amanda Askell},
  \bibinfo{person}{Sandhini Agarwal}, \bibinfo{person}{Ariel Herbert-Voss},
  \bibinfo{person}{Gretchen Krueger}, \bibinfo{person}{Tom Henighan},
  \bibinfo{person}{Rewon Child}, \bibinfo{person}{Aditya Ramesh},
  \bibinfo{person}{Daniel~M. Ziegler}, \bibinfo{person}{Jeffrey Wu},
  \bibinfo{person}{Clemens Winter}, \bibinfo{person}{Christopher Hesse},
  \bibinfo{person}{Mark Chen}, \bibinfo{person}{Eric Sigler},
  \bibinfo{person}{Mateusz Litwin}, \bibinfo{person}{Scott Gray},
  \bibinfo{person}{Benjamin Chess}, \bibinfo{person}{Jack Clark},
  \bibinfo{person}{Christopher Berner}, \bibinfo{person}{Sam McCandlish},
  \bibinfo{person}{Alec Radford}, \bibinfo{person}{Ilya Sutskever}, {and}
  \bibinfo{person}{Dario Amodei}.} \bibinfo{year}{2020}\natexlab{}.
\newblock \bibinfo{title}{Language Models are Few-Shot Learners}.
\newblock
\newblock
\newblock
\shownote{arXiv:2005.14165}.


\bibitem[Chen et~al\mbox{.}(2017)]%
        {chen2017jaid}
\bibfield{author}{\bibinfo{person}{Liushan Chen}, \bibinfo{person}{Yu Pei},
  {and} \bibinfo{person}{Carlo~A. Furia}.} \bibinfo{year}{2017}\natexlab{}.
\newblock \showarticletitle{Contract-based program repair without the
  contracts}. In \bibinfo{booktitle}{\emph{2017 32nd IEEE/ACM International
  Conference on Automated Software Engineering (ASE)}}.
  \bibinfo{pages}{637--647}.
\newblock


\bibitem[Chen et~al\mbox{.}(2021)]%
        {chen2021codex}
\bibfield{author}{\bibinfo{person}{Mark Chen}, \bibinfo{person}{Jerry Tworek},
  \bibinfo{person}{Heewoo Jun}, \bibinfo{person}{Qiming Yuan},
  \bibinfo{person}{Henrique~Ponde de Oliveira~Pinto}, \bibinfo{person}{Jared
  Kaplan}, \bibinfo{person}{Harri Edwards}, \bibinfo{person}{Yuri Burda},
  \bibinfo{person}{Nicholas Joseph}, \bibinfo{person}{Greg Brockman},
  \bibinfo{person}{Alex Ray}, \bibinfo{person}{Raul Puri},
  \bibinfo{person}{Gretchen Krueger}, \bibinfo{person}{Michael Petrov},
  \bibinfo{person}{Heidy Khlaaf}, \bibinfo{person}{Girish Sastry},
  \bibinfo{person}{Pamela Mishkin}, \bibinfo{person}{Brooke Chan},
  \bibinfo{person}{Scott Gray}, \bibinfo{person}{Nick Ryder},
  \bibinfo{person}{Mikhail Pavlov}, \bibinfo{person}{Alethea Power},
  \bibinfo{person}{Lukasz Kaiser}, \bibinfo{person}{Mohammad Bavarian},
  \bibinfo{person}{Clemens Winter}, \bibinfo{person}{Philippe Tillet},
  \bibinfo{person}{Felipe~Petroski Such}, \bibinfo{person}{Dave Cummings},
  \bibinfo{person}{Matthias Plappert}, \bibinfo{person}{Fotios Chantzis},
  \bibinfo{person}{Elizabeth Barnes}, \bibinfo{person}{Ariel Herbert-Voss},
  \bibinfo{person}{William~Hebgen Guss}, \bibinfo{person}{Alex Nichol},
  \bibinfo{person}{Alex Paino}, \bibinfo{person}{Nikolas Tezak},
  \bibinfo{person}{Jie Tang}, \bibinfo{person}{Igor Babuschkin},
  \bibinfo{person}{Suchir Balaji}, \bibinfo{person}{Shantanu Jain},
  \bibinfo{person}{William Saunders}, \bibinfo{person}{Christopher Hesse},
  \bibinfo{person}{Andrew~N. Carr}, \bibinfo{person}{Jan Leike},
  \bibinfo{person}{Josh Achiam}, \bibinfo{person}{Vedant Misra},
  \bibinfo{person}{Evan Morikawa}, \bibinfo{person}{Alec Radford},
  \bibinfo{person}{Matthew Knight}, \bibinfo{person}{Miles Brundage},
  \bibinfo{person}{Mira Murati}, \bibinfo{person}{Katie Mayer},
  \bibinfo{person}{Peter Welinder}, \bibinfo{person}{Bob McGrew},
  \bibinfo{person}{Dario Amodei}, \bibinfo{person}{Sam McCandlish},
  \bibinfo{person}{Ilya Sutskever}, {and} \bibinfo{person}{Wojciech Zaremba}.}
  \bibinfo{year}{2021}\natexlab{}.
\newblock \bibinfo{title}{Evaluating Large Language Models Trained on Code}.
\newblock
\newblock
\newblock
\shownote{arXiv:2107.03374}.


\bibitem[Chen et~al\mbox{.}(2019)]%
        {chen2018sequencer}
\bibfield{author}{\bibinfo{person}{Zimin Chen}, \bibinfo{person}{Steve
  Kommrusch}, \bibinfo{person}{Michele Tufano}, \bibinfo{person}{Louis-No{\"e}l
  Pouchet}, \bibinfo{person}{Denys Poshyvanyk}, {and} \bibinfo{person}{Martin
  Monperrus}.} \bibinfo{year}{2019}\natexlab{}.
\newblock \showarticletitle{SequenceR: Sequence-to-Sequence Learning for
  End-to-End Program Repair}.
\newblock \bibinfo{journal}{\emph{IEEE Transaction on Software Engineering}}
  (\bibinfo{year}{2019}).
\newblock


\bibitem[Chowdhery et~al\mbox{.}(2022)]%
        {chowdhery2022palm}
\bibfield{author}{\bibinfo{person}{Aakanksha Chowdhery},
  \bibinfo{person}{Sharan Narang}, \bibinfo{person}{Jacob Devlin},
  \bibinfo{person}{Maarten Bosma}, \bibinfo{person}{Gaurav Mishra},
  \bibinfo{person}{Adam Roberts}, \bibinfo{person}{Paul Barham},
  \bibinfo{person}{Hyung~Won Chung}, \bibinfo{person}{Charles Sutton},
  \bibinfo{person}{Sebastian Gehrmann}, \bibinfo{person}{Parker Schuh},
  \bibinfo{person}{Kensen Shi}, \bibinfo{person}{Sasha Tsvyashchenko},
  \bibinfo{person}{Joshua Maynez}, \bibinfo{person}{Abhishek Rao},
  \bibinfo{person}{Parker Barnes}, \bibinfo{person}{Yi Tay},
  \bibinfo{person}{Noam Shazeer}, \bibinfo{person}{Vinodkumar Prabhakaran},
  \bibinfo{person}{Emily Reif}, \bibinfo{person}{Nan Du}, \bibinfo{person}{Ben
  Hutchinson}, \bibinfo{person}{Reiner Pope}, \bibinfo{person}{James Bradbury},
  \bibinfo{person}{Jacob Austin}, \bibinfo{person}{Michael Isard},
  \bibinfo{person}{Guy Gur-Ari}, \bibinfo{person}{Pengcheng Yin},
  \bibinfo{person}{Toju Duke}, \bibinfo{person}{Anselm Levskaya},
  \bibinfo{person}{Sanjay Ghemawat}, \bibinfo{person}{Sunipa Dev},
  \bibinfo{person}{Henryk Michalewski}, \bibinfo{person}{Xavier Garcia},
  \bibinfo{person}{Vedant Misra}, \bibinfo{person}{Kevin Robinson},
  \bibinfo{person}{Liam Fedus}, \bibinfo{person}{Denny Zhou},
  \bibinfo{person}{Daphne Ippolito}, \bibinfo{person}{David Luan},
  \bibinfo{person}{Hyeontaek Lim}, \bibinfo{person}{Barret Zoph},
  \bibinfo{person}{Alexander Spiridonov}, \bibinfo{person}{Ryan Sepassi},
  \bibinfo{person}{David Dohan}, \bibinfo{person}{Shivani Agrawal},
  \bibinfo{person}{Mark Omernick}, \bibinfo{person}{Andrew~M. Dai},
  \bibinfo{person}{Thanumalayan~Sankaranarayana Pillai}, \bibinfo{person}{Marie
  Pellat}, \bibinfo{person}{Aitor Lewkowycz}, \bibinfo{person}{Erica Moreira},
  \bibinfo{person}{Rewon Child}, \bibinfo{person}{Oleksandr Polozov},
  \bibinfo{person}{Katherine Lee}, \bibinfo{person}{Zongwei Zhou},
  \bibinfo{person}{Xuezhi Wang}, \bibinfo{person}{Brennan Saeta},
  \bibinfo{person}{Mark Diaz}, \bibinfo{person}{Orhan Firat},
  \bibinfo{person}{Michele Catasta}, \bibinfo{person}{Jason Wei},
  \bibinfo{person}{Kathy Meier-Hellstern}, \bibinfo{person}{Douglas Eck},
  \bibinfo{person}{Jeff Dean}, \bibinfo{person}{Slav Petrov}, {and}
  \bibinfo{person}{Noah Fiedel}.} \bibinfo{year}{2022}\natexlab{}.
\newblock \bibinfo{title}{PaLM: Scaling Language Modeling with Pathways}.
\newblock
\newblock
\showeprint[arxiv]{2204.02311}~[cs.CL]


\bibitem[Dallmeier and Zimmermann(2007)]%
        {dallmeier2007benchmark}
\bibfield{author}{\bibinfo{person}{Valentin Dallmeier} {and}
  \bibinfo{person}{Thomas Zimmermann}.} \bibinfo{year}{2007}\natexlab{}.
\newblock \showarticletitle{Extraction of Bug Localization Benchmarks from
  History}. In \bibinfo{booktitle}{\emph{Proceedings of the Twenty-Second
  IEEE/ACM International Conference on Automated Software Engineering}}
  (Atlanta, Georgia, USA) \emph{(\bibinfo{series}{ASE '07})}.
  \bibinfo{publisher}{Association for Computing Machinery},
  \bibinfo{address}{New York, NY, USA}, \bibinfo{pages}{433–436}.
\newblock
\showISBNx{9781595938824}


\bibitem[DeMarco et~al\mbox{.}(2014)]%
        {demacro2014nopol}
\bibfield{author}{\bibinfo{person}{Favio DeMarco}, \bibinfo{person}{Jifeng
  Xuan}, \bibinfo{person}{Daniel Le~Berre}, {and} \bibinfo{person}{Martin
  Monperrus}.} \bibinfo{year}{2014}\natexlab{}.
\newblock \showarticletitle{Automatic Repair of Buggy If Conditions and Missing
  Preconditions with SMT}. In \bibinfo{booktitle}{\emph{Proceedings of the 6th
  International Workshop on Constraints in Software Testing, Verification, and
  Analysis}} (Hyderabad, India) \emph{(\bibinfo{series}{CSTVA 2014})}.
  \bibinfo{pages}{30–39}.
\newblock
\showISBNx{9781450328470}


\bibitem[Devlin et~al\mbox{.}(2018)]%
        {devlin2018bert}
\bibfield{author}{\bibinfo{person}{Jacob Devlin}, \bibinfo{person}{Ming-Wei
  Chang}, \bibinfo{person}{Kenton Lee}, {and} \bibinfo{person}{Kristina
  Toutanova}.} \bibinfo{year}{2018}\natexlab{}.
\newblock \bibinfo{title}{BERT: Pre-training of Deep Bidirectional Transformers
  for Language Understanding}.
\newblock
\newblock
\newblock
\shownote{arXiv:1810.04805}.


\bibitem[Feng et~al\mbox{.}(2020)]%
        {feng2020codebert}
\bibfield{author}{\bibinfo{person}{Zhangyin Feng}, \bibinfo{person}{Daya Guo},
  \bibinfo{person}{Duyu Tang}, \bibinfo{person}{Nan Duan},
  \bibinfo{person}{Xiaocheng Feng}, \bibinfo{person}{Ming Gong},
  \bibinfo{person}{Linjun Shou}, \bibinfo{person}{Bing Qin},
  \bibinfo{person}{Ting Liu}, \bibinfo{person}{Daxin Jiang}, {and}
  \bibinfo{person}{Ming Zhou}.} \bibinfo{year}{2020}\natexlab{}.
\newblock \bibinfo{title}{CodeBERT: A Pre-Trained Model for Programming and
  Natural Languages}.
\newblock
\newblock
\newblock
\shownote{arXiv:2002.08155}.


\bibitem[Fried et~al\mbox{.}(2022)]%
        {fried2022incoder}
\bibfield{author}{\bibinfo{person}{Daniel Fried}, \bibinfo{person}{Armen
  Aghajanyan}, \bibinfo{person}{Jessy Lin}, \bibinfo{person}{Sida Wang},
  \bibinfo{person}{Eric Wallace}, \bibinfo{person}{Freda Shi},
  \bibinfo{person}{Ruiqi Zhong}, \bibinfo{person}{Wen-tau Yih},
  \bibinfo{person}{Luke Zettlemoyer}, {and} \bibinfo{person}{Mike Lewis}.}
  \bibinfo{year}{2022}\natexlab{}.
\newblock \bibinfo{title}{InCoder: A Generative Model for Code Infilling and
  Synthesis}.
\newblock
\newblock
\newblock
\shownote{arXiv:2204.05999}.


\bibitem[Ghanbari et~al\mbox{.}(2019)]%
        {ghanbari2019prapr}
\bibfield{author}{\bibinfo{person}{Ali Ghanbari}, \bibinfo{person}{Samuel
  Benton}, {and} \bibinfo{person}{Lingming Zhang}.}
  \bibinfo{year}{2019}\natexlab{}.
\newblock \showarticletitle{Practical Program Repair via Bytecode Mutation}. In
  \bibinfo{booktitle}{\emph{Proceedings of the 28th ACM SIGSOFT International
  Symposium on Software Testing and Analysis}} (Beijing, China)
  \emph{(\bibinfo{series}{ISSTA 2019})}. \bibinfo{publisher}{ACM},
  \bibinfo{pages}{19--30}.
\newblock
\showISBNx{978-1-4503-6224-5}


\bibitem[Guo et~al\mbox{.}(2021)]%
        {guo2021graphcodebert}
\bibfield{author}{\bibinfo{person}{Daya Guo}, \bibinfo{person}{Shuo Ren},
  \bibinfo{person}{Shuai Lu}, \bibinfo{person}{Zhangyin Feng},
  \bibinfo{person}{Duyu Tang}, \bibinfo{person}{Shujie Liu},
  \bibinfo{person}{Long Zhou}, \bibinfo{person}{Nan Duan},
  \bibinfo{person}{Alexey Svyatkovskiy}, \bibinfo{person}{Shengyu Fu},
  \bibinfo{person}{Michele Tufano}, \bibinfo{person}{Shao~Kun Deng},
  \bibinfo{person}{Colin Clement}, \bibinfo{person}{Dawn Drain},
  \bibinfo{person}{Neel Sundaresan}, \bibinfo{person}{Jian Yin},
  \bibinfo{person}{Daxin Jiang}, {and} \bibinfo{person}{Ming Zhou}.}
  \bibinfo{year}{2021}\natexlab{}.
\newblock \bibinfo{title}{GraphCodeBERT: Pre-training Code Representations with
  Data Flow}.
\newblock
\newblock
\showeprint[arxiv]{2009.08366}~[cs.SE]


\bibitem[Hua et~al\mbox{.}(2018)]%
        {hua2018sketchfix}
\bibfield{author}{\bibinfo{person}{Jinru Hua}, \bibinfo{person}{Mengshi Zhang},
  \bibinfo{person}{Kaiyuan Wang}, {and} \bibinfo{person}{Sarfraz Khurshid}.}
  \bibinfo{year}{2018}\natexlab{}.
\newblock \showarticletitle{SketchFix: A Tool for Automated Program Repair
  Approach Using Lazy Candidate Generation}. In
  \bibinfo{booktitle}{\emph{Proceedings of the 2018 26th ACM Joint Meeting on
  European Software Engineering Conference and Symposium on the Foundations of
  Software Engineering}} (Lake Buena Vista, FL, USA)
  \emph{(\bibinfo{series}{ESEC/FSE 2018})}. \bibinfo{publisher}{{ACM}},
  \bibinfo{pages}{888–891}.
\newblock
\showISBNx{9781450355735}


\bibitem[HuggingFace(2022)]%
        {HuggingFaceWebPage}
HuggingFace \bibinfo{year}{2022}\natexlab{}.
\newblock \bibinfo{title}{Hugging Face}.
\newblock
\newblock
\newblock
\shownote{\url{https://huggingface.co}}.


\bibitem[Husain et~al\mbox{.}(2020)]%
        {husain2020codesearchnet}
\bibfield{author}{\bibinfo{person}{Hamel Husain}, \bibinfo{person}{Ho-Hsiang
  Wu}, \bibinfo{person}{Tiferet Gazit}, \bibinfo{person}{Miltiadis Allamanis},
  {and} \bibinfo{person}{Marc Brockschmidt}.} \bibinfo{year}{2020}\natexlab{}.
\newblock \bibinfo{title}{CodeSearchNet Challenge: Evaluating the State of
  Semantic Code Search}.
\newblock
\newblock
\newblock
\shownote{arXiv:1909.09436}.


\bibitem[Jaro(1989)]%
        {jaro1989advances}
\bibfield{author}{\bibinfo{person}{Matthew~A Jaro}.}
  \bibinfo{year}{1989}\natexlab{}.
\newblock \showarticletitle{Advances in record-linkage methodology as applied
  to matching the 1985 census of Tampa, Florida}.
\newblock \bibinfo{journal}{\emph{J. Amer. Statist. Assoc.}}
  \bibinfo{volume}{84}, \bibinfo{number}{406} (\bibinfo{year}{1989}),
  \bibinfo{pages}{414--420}.
\newblock


\bibitem[JavaParser(2023)]%
        {javaparser}
JavaParser \bibinfo{year}{2023}\natexlab{}.
\newblock \bibinfo{title}{JavaParser}.
\newblock
\newblock
\newblock
\shownote{\url{https://javaparser.org}}.


\bibitem[Jiang et~al\mbox{.}(2019)]%
        {jiang2019infer}
\bibfield{author}{\bibinfo{person}{Jiajun Jiang}, \bibinfo{person}{Luyao Ren},
  \bibinfo{person}{Yingfei Xiong}, {and} \bibinfo{person}{Lingming Zhang}.}
  \bibinfo{year}{2019}\natexlab{}.
\newblock \showarticletitle{Inferring Program Transformations From Singular
  Examples via Big Code}. In \bibinfo{booktitle}{\emph{2019 34th IEEE/ACM
  International Conference on Automated Software Engineering (ASE)}}.
  \bibinfo{pages}{255--266}.
\newblock


\bibitem[Jiang et~al\mbox{.}(2018)]%
        {jiang2018simfix}
\bibfield{author}{\bibinfo{person}{Jiajun Jiang}, \bibinfo{person}{Yingfei
  Xiong}, \bibinfo{person}{Hongyu Zhang}, \bibinfo{person}{Qing Gao}, {and}
  \bibinfo{person}{Xiangqun Chen}.} \bibinfo{year}{2018}\natexlab{}.
\newblock \showarticletitle{Shaping program repair space with existing patches
  and similar code}. In \bibinfo{booktitle}{\emph{Proceedings of the 27th {ACM}
  {SIGSOFT} International Symposium on Software Testing and Analysis, {ISSTA}
  2018, Amsterdam, The Netherlands, July 16-21, 2018}},
  \bibfield{editor}{\bibinfo{person}{Frank Tip} {and} \bibinfo{person}{Eric
  Bodden}} (Eds.). \bibinfo{publisher}{{ACM}}, \bibinfo{pages}{298--309}.
\newblock


\bibitem[Jiang et~al\mbox{.}(2021)]%
        {jiang2021cure}
\bibfield{author}{\bibinfo{person}{Nan Jiang}, \bibinfo{person}{Thibaud
  Lutellier}, {and} \bibinfo{person}{Lin Tan}.}
  \bibinfo{year}{2021}\natexlab{}.
\newblock \showarticletitle{CURE: Code-Aware Neural Machine Translation for
  Automatic Program Repair}.
\newblock \bibinfo{journal}{\emph{2021 IEEE/ACM 43rd International Conference
  on Software Engineering (ICSE)}} (\bibinfo{date}{May} \bibinfo{year}{2021}).
\newblock


\bibitem[Just et~al\mbox{.}(2014)]%
        {just2014dfj}
\bibfield{author}{\bibinfo{person}{Ren\'{e} Just}, \bibinfo{person}{Darioush
  Jalali}, {and} \bibinfo{person}{Michael~D. Ernst}.}
  \bibinfo{year}{2014}\natexlab{}.
\newblock \showarticletitle{Defects4J: A Database of Existing Faults to Enable
  Controlled Testing Studies for Java Programs} \emph{(\bibinfo{series}{ISSTA
  2014})}. \bibinfo{publisher}{Association for Computing Machinery},
  \bibinfo{address}{New York, NY, USA}, \bibinfo{pages}{437–440}.
\newblock
\showISBNx{9781450326452}


\bibitem[Kanade et~al\mbox{.}(2020)]%
        {kanade2020cubert}
\bibfield{author}{\bibinfo{person}{Aditya Kanade}, \bibinfo{person}{Petros
  Maniatis}, \bibinfo{person}{Gogul Balakrishnan}, {and}
  \bibinfo{person}{Kensen Shi}.} \bibinfo{year}{2020}\natexlab{}.
\newblock \bibinfo{title}{Learning and Evaluating Contextual Embedding of
  Source Code}.
\newblock
\newblock


\bibitem[Kingma and Ba(2014)]%
        {ba2014adam}
\bibfield{author}{\bibinfo{person}{Diederik~P. Kingma} {and}
  \bibinfo{person}{Jimmy Ba}.} \bibinfo{year}{2014}\natexlab{}.
\newblock \bibinfo{title}{Adam: A Method for Stochastic Optimization}.
\newblock
\newblock
\showeprint[arxiv]{1412.6980}~[cs.CL]


\bibitem[Kolak et~al\mbox{.}(2022)]%
        {kolak2022patch}
\bibfield{author}{\bibinfo{person}{Sophia~D Kolak}, \bibinfo{person}{Ruben
  Martins}, \bibinfo{person}{Claire~Le Goues}, {and}
  \bibinfo{person}{Vincent~Josua Hellendoorn}.}
  \bibinfo{year}{2022}\natexlab{}.
\newblock \showarticletitle{Patch Generation with Language Models: Feasibility
  and Scaling Behavior}. In \bibinfo{booktitle}{\emph{Deep Learning for Code
  Workshop}}.
\newblock


\bibitem[Koyuncu et~al\mbox{.}(2020)]%
        {koyuncu2020fixminder}
\bibfield{author}{\bibinfo{person}{Anil Koyuncu}, \bibinfo{person}{Kui Liu},
  \bibinfo{person}{Tegawend{\'{e}}~F. Bissyand{\'{e}}},
  \bibinfo{person}{Dongsun Kim}, \bibinfo{person}{Jacques Klein},
  \bibinfo{person}{Martin Monperrus}, {and} \bibinfo{person}{Yves~Le Traon}.}
  \bibinfo{year}{2020}\natexlab{}.
\newblock \showarticletitle{FixMiner: Mining relevant fix patterns for
  automated program repair}.
\newblock \bibinfo{journal}{\emph{Empir. Softw. Eng.}} \bibinfo{volume}{25},
  \bibinfo{number}{3} (\bibinfo{year}{2020}), \bibinfo{pages}{1980--2024}.
\newblock


\bibitem[Kuznia et~al\mbox{.}(2022)]%
        {kuznia2022summ}
\bibfield{author}{\bibinfo{person}{Kirby Kuznia}, \bibinfo{person}{Swaroop
  Mishra}, \bibinfo{person}{Mihir Parmar}, {and} \bibinfo{person}{Chitta
  Baral}.} \bibinfo{year}{2022}\natexlab{}.
\newblock \bibinfo{title}{Less is More: Summary of Long Instructions is Better
  for Program Synthesis}.
\newblock
\newblock
\newblock
\shownote{arXiv:2203.08597}.


\bibitem[Le et~al\mbox{.}(2017)]%
        {le2017s3}
\bibfield{author}{\bibinfo{person}{Xuan-Bach~D Le}, \bibinfo{person}{Duc-Hiep
  Chu}, \bibinfo{person}{David Lo}, \bibinfo{person}{Claire Le~Goues}, {and}
  \bibinfo{person}{Willem Visser}.} \bibinfo{year}{2017}\natexlab{}.
\newblock \showarticletitle{S3: syntax-and semantic-guided repair synthesis via
  programming by examples}. In \bibinfo{booktitle}{\emph{Proceedings of the
  2017 11th Joint Meeting on Foundations of Software Engineering}}.
  \bibinfo{pages}{593--604}.
\newblock


\bibitem[Le et~al\mbox{.}(2016)]%
        {le2016hdrepair}
\bibfield{author}{\bibinfo{person}{Xuan Bach~D. Le}, \bibinfo{person}{David
  Lo}, {and} \bibinfo{person}{Claire Le~Goues}.}
  \bibinfo{year}{2016}\natexlab{}.
\newblock \showarticletitle{History Driven Program Repair}. In
  \bibinfo{booktitle}{\emph{2016 IEEE 23rd International Conference on Software
  Analysis, Evolution, and Reengineering (SANER)}}, Vol.~\bibinfo{volume}{1}.
  \bibinfo{pages}{213--224}.
\newblock


\bibitem[Le~Goues et~al\mbox{.}(2012)]%
        {legoues2012genprog}
\bibfield{author}{\bibinfo{person}{Claire Le~Goues}, \bibinfo{person}{ThanhVu
  Nguyen}, \bibinfo{person}{Stephanie Forrest}, {and} \bibinfo{person}{Westley
  Weimer}.} \bibinfo{year}{2012}\natexlab{}.
\newblock \showarticletitle{GenProg: A Generic Method for Automatic Software
  Repair}.
\newblock \bibinfo{journal}{\emph{IEEE Transactions on Software Engineering}}
  \bibinfo{volume}{38}, \bibinfo{number}{1} (\bibinfo{year}{2012}),
  \bibinfo{pages}{54--72}.
\newblock


\bibitem[Levenshtein et~al\mbox{.}(1966)]%
        {levenshtein1966binary}
\bibfield{author}{\bibinfo{person}{Vladimir~I Levenshtein} {et~al\mbox{.}}}
  \bibinfo{year}{1966}\natexlab{}.
\newblock \showarticletitle{Binary codes capable of correcting deletions,
  insertions, and reversals}. In \bibinfo{booktitle}{\emph{Soviet physics
  doklady}}, Vol.~\bibinfo{volume}{10}. Soviet Union,
  \bibinfo{pages}{707--710}.
\newblock


\bibitem[Lewis et~al\mbox{.}(2019)]%
        {lewis2019bart}
\bibfield{author}{\bibinfo{person}{Mike Lewis}, \bibinfo{person}{Yinhan Liu},
  \bibinfo{person}{Naman Goyal}, \bibinfo{person}{Marjan Ghazvininejad},
  \bibinfo{person}{Abdelrahman Mohamed}, \bibinfo{person}{Omer Levy},
  \bibinfo{person}{Ves Stoyanov}, {and} \bibinfo{person}{Luke Zettlemoyer}.}
  \bibinfo{year}{2019}\natexlab{}.
\newblock \bibinfo{title}{BART: Denoising Sequence-to-Sequence Pre-training for
  Natural Language Generation, Translation, and Comprehension}.
\newblock
\newblock
\newblock
\shownote{arXiv:1910.13461}.


\bibitem[Li et~al\mbox{.}(2019)]%
        {li2019deepfl}
\bibfield{author}{\bibinfo{person}{Xia Li}, \bibinfo{person}{Wei Li},
  \bibinfo{person}{Yuqun Zhang}, {and} \bibinfo{person}{Lingming Zhang}.}
  \bibinfo{year}{2019}\natexlab{}.
\newblock \showarticletitle{Deepfl: Integrating multiple fault diagnosis
  dimensions for deep fault localization}. In
  \bibinfo{booktitle}{\emph{Proceedings of the 28th ACM SIGSOFT International
  Symposium on Software Testing and Analysis}}. \bibinfo{pages}{169--180}.
\newblock


\bibitem[Li and Zhang(2017)]%
        {li2017transforming}
\bibfield{author}{\bibinfo{person}{Xia Li} {and} \bibinfo{person}{Lingming
  Zhang}.} \bibinfo{year}{2017}\natexlab{}.
\newblock \showarticletitle{Transforming programs and tests in tandem for fault
  localization}.
\newblock \bibinfo{journal}{\emph{Proceedings of the ACM on Programming
  Languages}} \bibinfo{volume}{1}, \bibinfo{number}{OOPSLA}
  (\bibinfo{year}{2017}), \bibinfo{pages}{1--30}.
\newblock


\bibitem[Li et~al\mbox{.}(2020)]%
        {li2020dlfix}
\bibfield{author}{\bibinfo{person}{Yi Li}, \bibinfo{person}{Shaohua Wang},
  {and} \bibinfo{person}{Tien~N. Nguyen}.} \bibinfo{year}{2020}\natexlab{}.
\newblock \showarticletitle{DLFix: Context-Based Code Transformation Learning
  for Automated Program Repair}. In \bibinfo{booktitle}{\emph{Proceedings of
  the ACM/IEEE 42nd International Conference on Software Engineering}} (Seoul,
  South Korea) \emph{(\bibinfo{series}{ICSE '20})}.
  \bibinfo{publisher}{Association for Computing Machinery},
  \bibinfo{address}{New York, NY, USA}, \bibinfo{pages}{602–614}.
\newblock
\showISBNx{9781450371216}


\bibitem[Liu et~al\mbox{.}(2019a)]%
        {liu2019tbar}
\bibfield{author}{\bibinfo{person}{Kui Liu}, \bibinfo{person}{Anil Koyuncu},
  \bibinfo{person}{Dongsun Kim}, {and} \bibinfo{person}{Tegawend\'{e}~F.
  Bissyand\'{e}}.} \bibinfo{year}{2019}\natexlab{a}.
\newblock \showarticletitle{TBar: Revisiting Template-Based Automated Program
  Repair}. In \bibinfo{booktitle}{\emph{Proceedings of the 28th ACM SIGSOFT
  International Symposium on Software Testing and Analysis}}
  \emph{(\bibinfo{series}{ISSTA 2019})}. \bibinfo{publisher}{ACM},
  \bibinfo{address}{New York, NY, USA}, \bibinfo{pages}{31–42}.
\newblock
\showISBNx{9781450362245}


\bibitem[Liu et~al\mbox{.}(2019b)]%
        {liu2019avatar}
\bibfield{author}{\bibinfo{person}{Kui Liu}, \bibinfo{person}{Anil Koyuncu},
  \bibinfo{person}{Dongsun Kim}, {and} \bibinfo{person}{Tegawend{\'e}
  F.~Bissyand{\'e}}.} \bibinfo{year}{2019}\natexlab{b}.
\newblock \showarticletitle{{AVATAR:} Fixing Semantic Bugs with Fix Patterns of
  Static Analysis Violations}. In \bibinfo{booktitle}{\emph{Proceedings of the
  26th IEEE International Conference on Software Analysis, Evolution, and
  Reengineering}}. IEEE, \bibinfo{pages}{456--467}.
\newblock


\bibitem[Long and Rinard(2015)]%
        {long2015spr}
\bibfield{author}{\bibinfo{person}{Fan Long} {and} \bibinfo{person}{Martin
  Rinard}.} \bibinfo{year}{2015}\natexlab{}.
\newblock \showarticletitle{Staged Program Repair with Condition Synthesis}. In
  \bibinfo{booktitle}{\emph{Proceedings of the 2015 10th Joint Meeting on
  Foundations of Software Engineering}} (Bergamo, Italy)
  \emph{(\bibinfo{series}{ESEC/FSE 2015})}. \bibinfo{address}{New York, NY,
  USA}, \bibinfo{pages}{166–178}.
\newblock


\bibitem[Long and Rinard(2016)]%
        {long2016prophet}
\bibfield{author}{\bibinfo{person}{Fan Long} {and} \bibinfo{person}{Martin
  Rinard}.} \bibinfo{year}{2016}\natexlab{}.
\newblock \showarticletitle{Automatic Patch Generation by Learning Correct
  Code}.
\newblock \bibinfo{journal}{\emph{SIGPLAN Not.}} \bibinfo{volume}{51},
  \bibinfo{number}{1} (\bibinfo{date}{jan} \bibinfo{year}{2016}),
  \bibinfo{pages}{298–312}.
\newblock
\showISSN{0362-1340}


\bibitem[Lou et~al\mbox{.}(2020)]%
        {lou2020can}
\bibfield{author}{\bibinfo{person}{Yiling Lou}, \bibinfo{person}{Ali Ghanbari},
  \bibinfo{person}{Xia Li}, \bibinfo{person}{Lingming Zhang},
  \bibinfo{person}{Haotian Zhang}, \bibinfo{person}{Dan Hao}, {and}
  \bibinfo{person}{Lu Zhang}.} \bibinfo{year}{2020}\natexlab{}.
\newblock \showarticletitle{Can automated program repair refine fault
  localization? a unified debugging approach}. In
  \bibinfo{booktitle}{\emph{Proceedings of the 29th ACM SIGSOFT International
  Symposium on Software Testing and Analysis}}. \bibinfo{pages}{75--87}.
\newblock


\bibitem[Lu et~al\mbox{.}(2021)]%
        {lu2021codexglue}
\bibfield{author}{\bibinfo{person}{Shuai Lu}, \bibinfo{person}{Daya Guo},
  \bibinfo{person}{Shuo Ren}, \bibinfo{person}{Junjie Huang},
  \bibinfo{person}{Alexey Svyatkovskiy}, \bibinfo{person}{Ambrosio Blanco},
  \bibinfo{person}{Colin Clement}, \bibinfo{person}{Dawn Drain},
  \bibinfo{person}{Daxin Jiang}, \bibinfo{person}{Duyu Tang},
  \bibinfo{person}{Ge Li}, \bibinfo{person}{Lidong Zhou},
  \bibinfo{person}{Linjun Shou}, \bibinfo{person}{Long Zhou},
  \bibinfo{person}{Michele Tufano}, \bibinfo{person}{Ming Gong},
  \bibinfo{person}{Ming Zhou}, \bibinfo{person}{Nan Duan},
  \bibinfo{person}{Neel Sundaresan}, \bibinfo{person}{Shao~Kun Deng},
  \bibinfo{person}{Shengyu Fu}, {and} \bibinfo{person}{Shujie Liu}.}
  \bibinfo{year}{2021}\natexlab{}.
\newblock \bibinfo{title}{CodeXGLUE: A Machine Learning Benchmark Dataset for
  Code Understanding and Generation}.
\newblock
\newblock
\showeprint[arxiv]{2102.04664}~[cs.SE]


\bibitem[Lutellier et~al\mbox{.}(2020)]%
        {lutellier2020coconut}
\bibfield{author}{\bibinfo{person}{Thibaud Lutellier},
  \bibinfo{person}{Hung~Viet Pham}, \bibinfo{person}{Lawrence Pang},
  \bibinfo{person}{Yitong Li}, \bibinfo{person}{Moshi Wei}, {and}
  \bibinfo{person}{Lin Tan}.} \bibinfo{year}{2020}\natexlab{}.
\newblock \showarticletitle{CoCoNuT: Combining Context-Aware Neural Translation
  Models Using Ensemble for Program Repair}. In
  \bibinfo{booktitle}{\emph{Proceedings of the 29th ACM SIGSOFT International
  Symposium on Software Testing and Analysis}} (Virtual Event, USA)
  \emph{(\bibinfo{series}{ISSTA 2020})}. \bibinfo{publisher}{Association for
  Computing Machinery}, \bibinfo{address}{New York, NY, USA},
  \bibinfo{pages}{101–114}.
\newblock
\showISBNx{9781450380089}


\bibitem[Martinez et~al\mbox{.}(2015)]%
        {martinez2015automatic}
\bibfield{author}{\bibinfo{person}{Matias Martinez}, \bibinfo{person}{Thomas
  Durieux}, \bibinfo{person}{Jifeng Xuan}, \bibinfo{person}{Romain Sommerard},
  {and} \bibinfo{person}{Martin Monperrus}.} \bibinfo{year}{2015}\natexlab{}.
\newblock \bibinfo{title}{Automatic Repair of Real Bugs: An Experience Report
  on the Defects4J Dataset}.
\newblock
\newblock
\newblock
\shownote{arXiv:1505.07002}.


\bibitem[Martinez and Monperrus(2016)]%
        {martinez2016astor}
\bibfield{author}{\bibinfo{person}{Matias Martinez} {and}
  \bibinfo{person}{Martin Monperrus}.} \bibinfo{year}{2016}\natexlab{}.
\newblock \showarticletitle{ASTOR: A Program Repair Library for Java (Demo)}.
  In \bibinfo{booktitle}{\emph{Proceedings of the 25th International Symposium
  on Software Testing and Analysis}} (Saarbr\"{u}cken, Germany)
  \emph{(\bibinfo{series}{ISSTA 2016})}. \bibinfo{publisher}{Association for
  Computing Machinery}, \bibinfo{address}{New York, NY, USA},
  \bibinfo{pages}{441–444}.
\newblock
\showISBNx{9781450343909}


\bibitem[Matteson(2018)]%
        {bug_loss}
\bibfield{author}{\bibinfo{person}{Scott Matteson}.}
  \bibinfo{year}{2018}\natexlab{}.
\newblock \showarticletitle{Report: Software failure caused \$1.7 trillion in
  financial losses in 2017}.
\newblock \bibinfo{journal}{\emph{TechRepublic}} (\bibinfo{year}{2018}).
\newblock
\newblock
\shownote{\url{https://www.techrepublic.com/article/report-software-failure-caused-1-7-trillion-in-financial-losses-in-2017/}}.


\bibitem[Mechtaev et~al\mbox{.}(2016)]%
        {mechtaev2016angelix}
\bibfield{author}{\bibinfo{person}{Sergey Mechtaev}, \bibinfo{person}{Jooyong
  Yi}, {and} \bibinfo{person}{Abhik Roychoudhury}.}
  \bibinfo{year}{2016}\natexlab{}.
\newblock \showarticletitle{Angelix: Scalable Multiline Program Patch Synthesis
  via Symbolic Analysis}. In \bibinfo{booktitle}{\emph{Proceedings of the 38th
  International Conference on Software Engineering}} (Austin, Texas)
  \emph{(\bibinfo{series}{ICSE '16})}. \bibinfo{pages}{691–701}.
\newblock


\bibitem[Nijkamp et~al\mbox{.}(2022)]%
        {Nijkamp2022CG}
\bibfield{author}{\bibinfo{person}{Erik Nijkamp}, \bibinfo{person}{Bo Pang},
  \bibinfo{person}{Hiroaki Hayashi}, \bibinfo{person}{Lifu Tu},
  \bibinfo{person}{Huan Wang}, \bibinfo{person}{Yingbo Zhou},
  \bibinfo{person}{Silvio Savarese}, {and} \bibinfo{person}{Caiming Xiong}.}
  \bibinfo{year}{2022}\natexlab{}.
\newblock \bibinfo{title}{CodeGen: An Open Large Language Model for Code with
  Multi-Turn Program Synthesis}.
\newblock
\newblock
\newblock
\shownote{arXiv:2203.13474}.


\bibitem[Omri and Sinz(2020)]%
        {omri2020defectpred}
\bibfield{author}{\bibinfo{person}{Safa Omri} {and} \bibinfo{person}{Carsten
  Sinz}.} \bibinfo{year}{2020}\natexlab{}.
\newblock \showarticletitle{Deep Learning for Software Defect Prediction: A
  Survey}. In \bibinfo{booktitle}{\emph{Proceedings of the IEEE/ACM 42nd
  International Conference on Software Engineering Workshops}} (Seoul, Republic
  of Korea) \emph{(\bibinfo{series}{ICSEW'20})}.
  \bibinfo{publisher}{Association for Computing Machinery},
  \bibinfo{address}{New York, NY, USA}, \bibinfo{pages}{209–214}.
\newblock


\bibitem[Papadakis and Le~Traon(2015)]%
        {papadakis2015metallaxis}
\bibfield{author}{\bibinfo{person}{Mike Papadakis} {and} \bibinfo{person}{Yves
  Le~Traon}.} \bibinfo{year}{2015}\natexlab{}.
\newblock \showarticletitle{Metallaxis-FL: mutation-based fault localization}.
\newblock \bibinfo{journal}{\emph{Software Testing, Verification and
  Reliability}} \bibinfo{volume}{25}, \bibinfo{number}{5-7}
  (\bibinfo{year}{2015}), \bibinfo{pages}{605--628}.
\newblock


\bibitem[Prenner et~al\mbox{.}(2022)]%
        {prenner2021codexws}
\bibfield{author}{\bibinfo{person}{Julian~Aron Prenner}, \bibinfo{person}{Hlib
  Babii}, {and} \bibinfo{person}{Romain Robbes}.}
  \bibinfo{year}{2022}\natexlab{}.
\newblock \showarticletitle{Can OpenAI's Codex Fix Bugs?: An evaluation on
  QuixBugs}. In \bibinfo{booktitle}{\emph{2022 IEEE/ACM International Workshop
  on Automated Program Repair (APR)}}. \bibinfo{pages}{69--75}.
\newblock


\bibitem[PyTorch(2022)]%
        {PyTorchWebPage}
PyTorch \bibinfo{year}{2022}\natexlab{}.
\newblock \bibinfo{title}{PyTorch}.
\newblock
\newblock
\newblock
\shownote{\url{http://pytorch.org}}.


\bibitem[Qi et~al\mbox{.}(2015)]%
        {qi2015gv}
\bibfield{author}{\bibinfo{person}{Zichao Qi}, \bibinfo{person}{Fan Long},
  \bibinfo{person}{Sara Achour}, {and} \bibinfo{person}{Martin Rinard}.}
  \bibinfo{year}{2015}\natexlab{}.
\newblock \showarticletitle{An Analysis of Patch Plausibility and Correctness
  for Generate-and-Validate Patch Generation Systems}. In
  \bibinfo{booktitle}{\emph{Proceedings of the 2015 International Symposium on
  Software Testing and Analysis}} (Baltimore, MD, USA)
  \emph{(\bibinfo{series}{ISSTA 2015})}. \bibinfo{publisher}{Association for
  Computing Machinery}, \bibinfo{address}{New York, NY, USA},
  \bibinfo{pages}{24–36}.
\newblock
\showISBNx{9781450336208}


\bibitem[Radford et~al\mbox{.}(2019)]%
        {radford2019gpt2}
\bibfield{author}{\bibinfo{person}{Alec Radford}, \bibinfo{person}{Jeff Wu},
  \bibinfo{person}{Rewon Child}, \bibinfo{person}{David Luan},
  \bibinfo{person}{Dario Amodei}, {and} \bibinfo{person}{Ilya Sutskever}.}
  \bibinfo{year}{2019}\natexlab{}.
\newblock \showarticletitle{Language Models are Unsupervised Multitask
  Learners}.
\newblock  (\bibinfo{year}{2019}).
\newblock


\bibitem[Raffel et~al\mbox{.}(2020)]%
        {raffel2020t5}
\bibfield{author}{\bibinfo{person}{Colin Raffel}, \bibinfo{person}{Noam
  Shazeer}, \bibinfo{person}{Adam Roberts}, \bibinfo{person}{Katherine Lee},
  \bibinfo{person}{Sharan Narang}, \bibinfo{person}{Michael Matena},
  \bibinfo{person}{Yanqi Zhou}, \bibinfo{person}{Wei Li}, {and}
  \bibinfo{person}{Peter~J. Liu}.} \bibinfo{year}{2020}\natexlab{}.
\newblock \showarticletitle{Exploring the Limits of Transfer Learning with a
  Unified Text-to-Text Transformer}.
\newblock \bibinfo{journal}{\emph{J. Mach. Learn. Res.}} (\bibinfo{date}{jan}
  \bibinfo{year}{2020}).
\newblock


\bibitem[Ratcliff and Metzener(1988)]%
        {ratcliff1988pattern}
\bibfield{author}{\bibinfo{person}{John~W Ratcliff} {and}
  \bibinfo{person}{David~E Metzener}.} \bibinfo{year}{1988}\natexlab{}.
\newblock \showarticletitle{Pattern-matching-the gestalt approach}.
\newblock \bibinfo{journal}{\emph{Dr Dobbs Journal}} \bibinfo{volume}{13},
  \bibinfo{number}{7} (\bibinfo{year}{1988}), \bibinfo{pages}{46}.
\newblock


\bibitem[Reimers and Gurevych(2019)]%
        {reimers2019sentence}
\bibfield{author}{\bibinfo{person}{Nils Reimers} {and} \bibinfo{person}{Iryna
  Gurevych}.} \bibinfo{year}{2019}\natexlab{}.
\newblock \showarticletitle{Sentence-bert: Sentence embeddings using siamese
  bert-networks}.
\newblock \bibinfo{journal}{\emph{arXiv preprint arXiv:1908.10084}}
  (\bibinfo{year}{2019}).
\newblock


\bibitem[Richards(1990)]%
        {bug_safety}
\bibfield{author}{\bibinfo{person}{Evelyn Richards}.}
  \bibinfo{year}{1990}\natexlab{}.
\newblock \showarticletitle{Software's Dangerous Aspect}.
\newblock \bibinfo{journal}{\emph{The Washington Post}} (\bibinfo{year}{1990}).
\newblock
\newblock
\shownote{\url{https://www.washingtonpost.com/archive/politics/1990/12/09/softwares-dangerous-aspect/9b2e9243-8deb-4ac7-9e8f-968de0806e5e/}}.


\bibitem[Robertson et~al\mbox{.}(2009)]%
        {robertson2009probabilistic}
\bibfield{author}{\bibinfo{person}{Stephen Robertson}, \bibinfo{person}{Hugo
  Zaragoza}, {et~al\mbox{.}}} \bibinfo{year}{2009}\natexlab{}.
\newblock \showarticletitle{The probabilistic relevance framework: BM25 and
  beyond}.
\newblock \bibinfo{journal}{\emph{Foundations and Trends{\textregistered} in
  Information Retrieval}} \bibinfo{volume}{3}, \bibinfo{number}{4}
  (\bibinfo{year}{2009}), \bibinfo{pages}{333--389}.
\newblock


\bibitem[Saha et~al\mbox{.}(2017)]%
        {saha2017elixir}
\bibfield{author}{\bibinfo{person}{Ripon~K. Saha}, \bibinfo{person}{Yingjun
  Lyu}, \bibinfo{person}{Hiroaki Yoshida}, {and} \bibinfo{person}{Mukul~R.
  Prasad}.} \bibinfo{year}{2017}\natexlab{}.
\newblock \showarticletitle{Elixir: Effective object-oriented program repair}.
  In \bibinfo{booktitle}{\emph{2017 32nd IEEE/ACM International Conference on
  Automated Software Engineering (ASE)}}. \bibinfo{pages}{648--659}.
\newblock


\bibitem[Sutskever et~al\mbox{.}(2014)]%
        {sutskever2014mt}
\bibfield{author}{\bibinfo{person}{Ilya Sutskever}, \bibinfo{person}{Oriol
  Vinyals}, {and} \bibinfo{person}{Quoc~V. Le}.}
  \bibinfo{year}{2014}\natexlab{}.
\newblock \bibinfo{title}{Sequence to Sequence Learning with Neural Networks}.
\newblock
\newblock
\newblock
\shownote{arXiv:1409.3215}.


\bibitem[Tufano et~al\mbox{.}(2018)]%
        {tufano2018empstudy}
\bibfield{author}{\bibinfo{person}{Michele Tufano}, \bibinfo{person}{Cody
  Watson}, \bibinfo{person}{Gabriele Bavota}, \bibinfo{person}{Massimiliano
  Di~Penta}, \bibinfo{person}{Martin White}, {and} \bibinfo{person}{Denys
  Poshyvanyk}.} \bibinfo{year}{2018}\natexlab{}.
\newblock \showarticletitle{An Empirical Investigation into Learning Bug-Fixing
  Patches in the Wild via Neural Machine Translation}. In
  \bibinfo{booktitle}{\emph{Proceedings of the 33rd ACM/IEEE International
  Conference on Automated Software Engineering}}. \bibinfo{pages}{832–837}.
\newblock


\bibitem[Tunstall et~al\mbox{.}(2022)]%
        {tunstall2022natural}
\bibfield{author}{\bibinfo{person}{Lewis Tunstall}, \bibinfo{person}{Leandro
  von Werra}, {and} \bibinfo{person}{Thomas Wolf}.}
  \bibinfo{year}{2022}\natexlab{}.
\newblock \bibinfo{booktitle}{\emph{Natural language processing with
  transformers}}.
\newblock \bibinfo{publisher}{" O'Reilly Media, Inc."}.
\newblock


\bibitem[Vaswani et~al\mbox{.}(2017)]%
        {vaswani2017attention}
\bibfield{author}{\bibinfo{person}{Ashish Vaswani}, \bibinfo{person}{Noam
  Shazeer}, \bibinfo{person}{Niki Parmar}, \bibinfo{person}{Jakob Uszkoreit},
  \bibinfo{person}{Llion Jones}, \bibinfo{person}{Aidan~N. Gomez},
  \bibinfo{person}{Lukasz Kaiser}, {and} \bibinfo{person}{Illia Polosukhin}.}
  \bibinfo{year}{2017}\natexlab{}.
\newblock \showarticletitle{Attention Is All You Need}.
\newblock  (\bibinfo{year}{2017}).
\newblock
\newblock
\shownote{arXiv:1706.03762}.


\bibitem[Wang et~al\mbox{.}(2022)]%
        {wang2022bridge}
\bibfield{author}{\bibinfo{person}{Deze Wang}, \bibinfo{person}{Zhouyang Jia},
  \bibinfo{person}{Shanshan Li}, \bibinfo{person}{Yue Yu}, \bibinfo{person}{Yun
  Xiong}, \bibinfo{person}{Wei Dong}, {and} \bibinfo{person}{Xiangke Liao}.}
  \bibinfo{year}{2022}\natexlab{}.
\newblock \showarticletitle{Bridging Pre-Trained Models and Downstream Tasks
  for Source Code Understanding}. In \bibinfo{booktitle}{\emph{Proceedings of
  the 44th International Conference on Software Engineering}} (Pittsburgh,
  Pennsylvania) \emph{(\bibinfo{series}{ICSE '22})}.
  \bibinfo{publisher}{Association for Computing Machinery},
  \bibinfo{address}{New York, NY, USA}, \bibinfo{pages}{287–298}.
\newblock


\bibitem[Wen et~al\mbox{.}(2018)]%
        {wen2018capgen}
\bibfield{author}{\bibinfo{person}{Ming Wen}, \bibinfo{person}{Junjie Chen},
  \bibinfo{person}{Rongxin Wu}, \bibinfo{person}{Dan Hao}, {and}
  \bibinfo{person}{Shing-Chi Cheung}.} \bibinfo{year}{2018}\natexlab{}.
\newblock \showarticletitle{Context-Aware Patch Generation for Better Automated
  Program Repair}. In \bibinfo{booktitle}{\emph{Proceedings of the 40th
  International Conference on Software Engineering}} (Gothenburg, Sweden)
  \emph{(\bibinfo{series}{ICSE '18})}. \bibinfo{pages}{1–11}.
\newblock
\showISBNx{9781450356381}


\bibitem[Wettig et~al\mbox{.}(2022)]%
        {wettig2022mask15}
\bibfield{author}{\bibinfo{person}{Alexander Wettig}, \bibinfo{person}{Tianyu
  Gao}, \bibinfo{person}{Zexuan Zhong}, {and} \bibinfo{person}{Danqi Chen}.}
  \bibinfo{year}{2022}\natexlab{}.
\newblock \bibinfo{title}{Should You Mask 15\% in Masked Language Modeling?}
\newblock
\newblock
\showeprint[arxiv]{2202.08005}~[cs.CL]


\bibitem[Winkler(1990)]%
        {winkler1990string}
\bibfield{author}{\bibinfo{person}{William~E Winkler}.}
  \bibinfo{year}{1990}\natexlab{}.
\newblock \showarticletitle{String comparator metrics and enhanced decision
  rules in the Fellegi-Sunter model of record linkage.}
\newblock  (\bibinfo{year}{1990}).
\newblock


\bibitem[Wong et~al\mbox{.}(2016)]%
        {wong2016fl}
\bibfield{author}{\bibinfo{person}{W.~Eric Wong}, \bibinfo{person}{Ruizhi Gao},
  \bibinfo{person}{Yihao Li}, \bibinfo{person}{Rui Abreu}, {and}
  \bibinfo{person}{Franz Wotawa}.} \bibinfo{year}{2016}\natexlab{}.
\newblock \showarticletitle{A Survey on Software Fault Localization}.
\newblock \bibinfo{journal}{\emph{IEEE Transactions on Software Engineering}}
  \bibinfo{volume}{42}, \bibinfo{number}{8} (\bibinfo{year}{2016}),
  \bibinfo{pages}{707--740}.
\newblock


\bibitem[Xia et~al\mbox{.}(2023)]%
        {xia2023repairstudy}
\bibfield{author}{\bibinfo{person}{Chunqiu~Steven Xia},
  \bibinfo{person}{Yuxiang Wei}, {and} \bibinfo{person}{Lingming Zhang}.}
  \bibinfo{year}{2023}\natexlab{}.
\newblock \showarticletitle{Automated Program Repair in the Era of Large
  Pre-trained Language Models}. In \bibinfo{booktitle}{\emph{Proceedings of the
  ACM/IEEE 45th International Conference on Software Engineering}}
  \emph{(\bibinfo{series}{ICSE '23})}.
\newblock


\bibitem[Xia and Zhang(2022)]%
        {xia2022alpharepair}
\bibfield{author}{\bibinfo{person}{Chunqiu~Steven Xia} {and}
  \bibinfo{person}{Lingming Zhang}.} \bibinfo{year}{2022}\natexlab{}.
\newblock \showarticletitle{Less Training, More Repairing Please: Revisiting
  Automated Program Repair via Zero-Shot Learning}. In
  \bibinfo{booktitle}{\emph{Proceedings of the 30th ACM Joint European Software
  Engineering Conference and Symposium on the Foundations of Software
  Engineering}} \emph{(\bibinfo{series}{ESEC/FSE 2022})}.
\newblock


\bibitem[Xu et~al\mbox{.}(2022)]%
        {xu2022systematic}
\bibfield{author}{\bibinfo{person}{Frank~F. Xu}, \bibinfo{person}{Uri Alon},
  \bibinfo{person}{Graham Neubig}, {and} \bibinfo{person}{Vincent~Josua
  Hellendoorn}.} \bibinfo{year}{2022}\natexlab{}.
\newblock \showarticletitle{A Systematic Evaluation of Large Language Models of
  Code}. In \bibinfo{booktitle}{\emph{Proceedings of the 6th ACM SIGPLAN
  International Symposium on Machine Programming}} (San Diego, CA, USA)
  \emph{(\bibinfo{series}{MAPS 2022})}. \bibinfo{publisher}{Association for
  Computing Machinery}, \bibinfo{address}{New York, NY, USA},
  \bibinfo{pages}{1–10}.
\newblock


\bibitem[Ye et~al\mbox{.}(2022b)]%
        {ye2022selfapr}
\bibfield{author}{\bibinfo{person}{He Ye}, \bibinfo{person}{Matias Martinez},
  \bibinfo{person}{Xiapu Luo}, \bibinfo{person}{Tao Zhang}, {and}
  \bibinfo{person}{Martin Monperrus}.} \bibinfo{year}{2022}\natexlab{b}.
\newblock \showarticletitle{SelfAPR: Self-supervised Program Repair with Test
  Execution Diagnostics}. In \bibinfo{booktitle}{\emph{37th IEEE/ACM
  International Conference on Automated Software Engineering}}
  \emph{(\bibinfo{series}{ASE22})}. \bibinfo{publisher}{Association for
  Computing Machinery}, Article \bibinfo{articleno}{92},
  \bibinfo{numpages}{13}~pages.
\newblock


\bibitem[Ye et~al\mbox{.}(2022a)]%
        {ye2022rewardrepair}
\bibfield{author}{\bibinfo{person}{He Ye}, \bibinfo{person}{Matias Martinez},
  {and} \bibinfo{person}{Martin Monperrus}.} \bibinfo{year}{2022}\natexlab{a}.
\newblock \showarticletitle{Neural Program Repair with Execution-based
  Backpropagation}. In \bibinfo{booktitle}{\emph{2022 IEEE/ACM 44th
  International Conference on Software Engineering (ICSE)}}.
  \bibinfo{pages}{1506--1518}.
\newblock


\bibitem[Yue~Wang and Hoi(2021)]%
        {wang2021codet5}
\bibfield{author}{\bibinfo{person}{Shafiq~Joty Yue~Wang, Weishi~Wang} {and}
  \bibinfo{person}{Steven~C.H. Hoi}.} \bibinfo{year}{2021}\natexlab{}.
\newblock \showarticletitle{CodeT5: Identifier-aware Unified Pre-trained
  Encoder-Decoder Models for Code Understanding and Generation}. In
  \bibinfo{booktitle}{\emph{Proceedings of the 2021 Conference on Empirical
  Methods in Natural Language Processing, EMNLP 2021}}.
\newblock


\bibitem[Zhang et~al\mbox{.}(2013)]%
        {zhang2013injecting}
\bibfield{author}{\bibinfo{person}{Lingming Zhang}, \bibinfo{person}{Lu Zhang},
  {and} \bibinfo{person}{Sarfraz Khurshid}.} \bibinfo{year}{2013}\natexlab{}.
\newblock \showarticletitle{Injecting mechanical faults to localize developer
  faults for evolving software}.
\newblock \bibinfo{journal}{\emph{ACM SIGPLAN Notices}} \bibinfo{volume}{48},
  \bibinfo{number}{10} (\bibinfo{year}{2013}), \bibinfo{pages}{765--784}.
\newblock


\bibitem[Zhu et~al\mbox{.}(2021)]%
        {zhu2021recoder}
\bibfield{author}{\bibinfo{person}{Qihao Zhu}, \bibinfo{person}{Zeyu Sun},
  \bibinfo{person}{Yuan-an Xiao}, \bibinfo{person}{Wenjie Zhang},
  \bibinfo{person}{Kang Yuan}, \bibinfo{person}{Yingfei Xiong}, {and}
  \bibinfo{person}{Lu Zhang}.} \bibinfo{year}{2021}\natexlab{}.
\newblock \showarticletitle{A Syntax-Guided Edit Decoder for Neural Program
  Repair}. In \bibinfo{booktitle}{\emph{Proceedings of the 29th ACM Joint
  Meeting on European Software Engineering Conference and Symposium on the
  Foundations of Software Engineering}}. \bibinfo{publisher}{ACM},
  \bibinfo{address}{New York, NY, USA}, \bibinfo{pages}{341–353}.
\newblock
\showISBNx{9781450385626}


\end{thebibliography}
